\documentclass{autart}
\usepackage{amsfonts,amssymb,latexsym,amsmath,amscd,theorem,epic,epsfig,graphics}
\usepackage{color,enumerate}
\newtheorem{theorem}{Theorem}
\newtheorem{remark}{Remark}
\newtheorem{lemma}{Lemma}
\newtheorem{proposition}{Proposition}

\newtheorem{corollary}{Corollary}

\newtheorem{question}{Question}
\newcommand{\tet}{\theta}

\newcommand{\bydef}{\stackrel{\Delta}{=}}

\newcommand{\bbZ}{{\mathbb Z}}
\newcommand{\bbN}{{\mathbb N}}
\newcommand{\bbR}{{\mathbb R}}

\DeclareMathOperator{\col}{col}

\begin{document}
\begin{frontmatter}

%%%%%%%%%%%%%%%%%%%%%%%%%%%%%%%%%%%%%%%%%%
%%%%%%%%%%%%%%%%%%%%%%%%%%%%%%%%%%%%%%%%%%%%%%%%%%%%%%%%%%%%%%%%%%%%%%%%%%

\title {Generalized Jensen Inequalities with Application to Stability Analysis of Systems with Distributed Delays over Infinite Time-Horizons  \thanksref{footnoteinfo}}
\thanks[footnoteinfo]{This work was partially supported by the National Natural Science Foundation of China (grant no. 61503026, 61440058), 
the Knut and Alice Wallenberg Foundation, the Swedish Research
Council, and the Israel Science Foundation (grant no. 754/10 and 1128/14).}
\author[First]{Kun Liu}
\author[Second]{Emilia Fridman}
\author[Third]{Karl Henrik Johansson}
\author[First]{Yuanqing Xia}
\address [First]{School of Automation,
Beijing Institute of Technology, 100081 Beijing, China (e-mails: kunliubit, xia$\_$yuanqing@bit.edu.cn).}
\address[Second]{School of Electrical Engineering, Tel Aviv University,
69978 Tel Aviv, Israel (e-mail: emilia@eng.tau.ac.il).}
\address[Third]{ACCESS Linnaeus Centre and School of Electrical Engineering,
KTH Royal Institute of Technology, SE-100 44 Stockholm, Sweden (e-mail: kallej@kth.se).}

\begin{abstract}
%In the last decade, the integral inequality has been intensively used in the context of time-delay
%systems since it is an appropriate tool to derive tractable stability conditions expressed in terms
%of linear matrix inequalities (LMIs).
%This paper presents some improved inequalities with infinite intervals of integration.
%Their application to stability analysis for time-delay system with Gamma-distributed delays is provided with a
%numerical example.
%
The Jensen inequality has been recognized as a powerful tool to deal with the stability of time-delay
systems. Recently, a new inequality that encompasses the Jensen inequality was proposed  for the stability analysis of systems with finite delays.
In this paper,
we first present a generalized integral inequality and its double integral extension. It is shown how these inequalities can be applied to improve the stability result for linear continuous-time systems with gamma-distributed delays.
Then, for the discrete-time counterpart we provide an extended Jensen summation inequality with infinite sequences, which leads to less conservative stability conditions for linear discrete-time systems with poisson-distributed delays.
The improvements obtained thanks to the introduced generalized inequalities are demonstrated by examples. % with gamma-distributed and poisson-distributed delays, respectively.
\end{abstract}

\end{frontmatter}

\section{Introduction}
%Time-delay systems are also called
%systems with aftereffect or dead-time, hereditary systems,
%equations with deviating argument, or differential-difference
%equations \cite{FridmanBook14}, \cite{RichardAut03}.
Time-delay often appears in many  control systems either in the state, the control input, or the measurements.
During the last two decades, the stability of time-delay systems has received
considerable attention (e.g., \cite{FridmanBook14}, \cite{GuBook03}, \cite{NicBook01}, \cite{RichardAut03} and references therein). One of the most popular approaches is the use of Lyapunov-Krasovskii functionals (LKF) to derive stability
conditions (e.g., \cite{AribaGouaisbautCDC08}, \cite{KunAut09}, \cite{KolRichardTAC1999}, \cite{XiaSCL03}). The choice of the Lyapunov functional and the method of bounding an integral term in the derivative of the LKF are important ways to reduce conservativeness of the stability results. %In  the existing literature on this topic,
The Jensen inequality \cite{GuBook03}, has been widely used as an efficient bounding technique, although %it is
at a price of an unavoidable conservativeness \cite{GuIJRNC03}, \cite{KunFrSCL14saturation}. The Jensen inequality claims that for any continuous function $\omega: [a, b] \to \bbR^n$ and $n \times n$ positive definite matrix~$R,$
\begin{equation}
\nonumber %\label{Jensen}
\begin{array}{ll}
\int_{a}^{b} \omega^T(s) R \omega(s) ds
\geq {1 \over b-a} \int_{a}^{b}\omega^T(s) ds R
\int_{a}^{b} \omega(s) ds
\end{array}
\end{equation}
holds.
There is a discrete counterpart, which involves
sums instead of integrals \cite{FridmanBook14}, \cite{FridmanIJC03}.

%\vspace{-.1cm}
Some recent efforts have been made to overcome the conservativeness induced by the Jensen inequality when applied to the stability
analysis of time-delay systems.
The bound on the gap of the Jensen
inequality was analyzed in \cite{BriatTAC11} by using
the Gr$\ddot {\rm u}$ss inequality.
%In \cite{BriatTAC11}, an upper-bound of the conservativeness induced by Jensen inequality
%is proposed by using Gr¡§uss inequality. This last result is
%then employed to derive equivalent new inequalities well
%fitted for an optimisation scheme purpose like LMIs.
%
Based on the Wirtinger inequality \cite{KunAut12},
Seuret and Gouaisbaut \cite{SeuretTDS12}
derived an extended integral inequality, which encompasses
Jensen inequality as a particular case.
Recently, the inequality they proposed was further refined
in \cite{SeuretAut13wirtinger}.
By combining the newly developed integral inequality and an augmented Lyapunov
functional, a remarkable result was obtained for systems with constant discrete
and distributed delays.
%
%the use of the Jensen inequality
%(see [Gu et al., 2003] or Appendix F.1) used to bound an integral term in the derivative
%of the Lyapunov-Krasovskii functional. It is also used in approaches based on Lyapunov-
%Krasovskii functionals as a efficient bounding technique
%
%
%
%
%
%
%
%
%The Jensen¡¯s Inequality [1] has had a tremendous impact on many
%different fields; e.g., convex analysis, probability theory, information
%theory, statistics, control and systems theory . application in time-delay systems [7], [3], [8],
%[4] concerns the bounding of integral quadratic terms of the form
%arising in some approaches based on Lyapunov¨C
%Krasovskii functionals (LKFs). Discrete counterparts, involving
%sums instead of integrals, also exist; see, e.g., [9], [10].
%The Jensen¡¯s Inequality has been widely used both in continuous and discrete time system.
%Jensen¡¯s inequality have been recognised as a powerful tool to obtain efficient results
%
%
%\subsection{Integral inequality with finite intervals of integration}
%In order to reduce the conservativeness of the Jensen inequality,
Let us recall the inequality provided  in \cite{SeuretAut13wirtinger} (see  \cite{SeuretTAC15discrete} for the discrete counterpart):
for any continuous function $\omega: [a, b] \to \bbR^n$ and $n \times n$ positive definite matrix $R,$ the inequality
%\vspace{-0.381cm}
\begin{equation}
\label{Inequality:Alex}
\begin{array}{ll}
\int_{a}^{b} \omega^T(s) R \omega(s) ds
&\geq {1 \over b-a} \int_{a}^{b}\omega^T(s) ds R
\int_{a}^{b} \omega(s) ds\\
&\ \ \ +{3 \over {b-a}} \Omega^T R \Omega
\end{array}
\end{equation}
holds, where
\vspace{-0.4cm}
\begin{equation}
\begin{array}{ll}
\label{Omega:Alex}
\Omega=\int_{a}^{b}\omega(s) ds-{2 \over b-a}\int_{a}^{b}\int_{a}^s \omega(r) dr ds.
\end{array}
\end{equation}
To prove (\ref{Inequality:Alex}), a function $f(u), \ u \in [a, b],$ was introduced in \cite{SeuretAut13wirtinger} as follows:
\begin{equation}\label{Inequality:Alex:f}
\begin{array}{ll}
f(u)=\dot z (u)=\omega(u)-{1 \over b-a} \int_{a}^{b}\omega(s) ds-{{a+b-2u} \over (b-a)^2} \Theta,
\end{array}
\end{equation}
where $\Theta \in \bbR^n$ is a constant vector to be defined and %as ${3 \over (b-a)^2} \Omega$ and
$z(u)=\int_{a}^{u}\omega(s)ds-{u-a \over b-a} \int_{a}^{b}\omega(s) ds-{(b-u)(u-a)\over (b-a)^2} \Theta,$ $u \in~[a, b]$.
It is noted that $\int_{a}^{b}(a+b-2u)du=0$ plays an important role in the utilization of (\ref{Inequality:Alex:f}).
Since $\Theta$ is a constant vector, it is obvious that in (\ref{Inequality:Alex:f}), $a+b-2u$ could be replaced by $c(a+b-2u), \ c \in \bbR \backslash \{0\},$ because $\int_{a}^{b}c(a+b-2u)du=0$.
By using a more general auxiliary function $g(u)$ with $\int_{a}^{b}g(u)du=0,$
an extended integral inequality, which included
the one proposed in~\cite{SeuretAut13wirtinger} as a particular case, was provided in \cite{ParkJFI2015}.

%Therefore, it is natural to address the following question:
%
%\begin{enumerate}
%\item[]
%Question 1: If we replace $a+b-2u$ in (\ref{Inequality:Alex:f}) by a more
%general scalar function $g(u)$ with
%$\int_{a}^{b}g(u)du=0$ and $g(u)$ not equivalent to $0$ for all $u \in [a, b]$, what is the corresponding result?
%\end{enumerate}
%

%\subsection{Integral inequalities with infinite intervals of integration}
%On the other hand, it was pointed out in \cite{Roesch05} that gamma-distributed delays with a gap also appear in the control over communication networks.
Recently, the stability analysis of systems with gamma-distributed delays was studied \cite{SolomonFridmanAut13}. The Lyapunov-based analysis was based on two kinds of integral inequalities with infinite intervals of
integration: given an $n \times n$ positive definite matrix $R,$ a scalar
$h \geq 0,$ a vector function
$\omega: [0, +\infty) \to \bbR^n$ and
a scalar function
$K: [0, +\infty) \to \bbR^{+}$
such that the integrations concerned are well
defined, the following inequalities
\begin{equation}
\label{Inequality:infinite:EF}
\begin{array}{ll}
\int_{0}^{+\infty}K(s) \omega^T(s) R \omega(s) ds\\
\geq K_0^{-1} \int_{0}^{+\infty}K(s)  \omega^T(s) ds  R
\int_{0}^{+\infty}K(s)  \omega(s) ds
\end{array}
\end{equation}
and
\begin{equation}
\label{Inequality:infinite:doub:EF}
\begin{array}{ll}
\int_{0}^{+\infty} \int_{t-\tet-h}^{t}K(\tet) \omega^T(s) R \omega(s) ds d \tet \\
\geq K_{1h}^{-1} \int_{0}^{+\infty} \int_{t-\tet-h}^{t} K(\tet)  \omega^T(s) ds d \tet R \\
\ \ \ \times \int_{0}^{+\infty}\int_{t-\tet-h}^{t} K(\tet)   \omega(s) ds  d \tet,
\end{array}
\end{equation}
hold, where $K_0=\int_{0}^{+\infty} K(s) ds$ and $K_{1h}=\int_{0}^{+\infty} K(s)  (s+h)ds$.
%\begin{equation} \nonumber
%\begin{array}{ll}
%K_0=\int_{0}^{+\infty} |K(s)| ds,\\
%K_{1h}=\int_{0}^{+\infty} |K(s)| (s+h)ds.
%\end{array}
%\end{equation}
%
The inequalities (\ref{Inequality:infinite:EF}) and (\ref{Inequality:infinite:doub:EF}) were used in~\cite{SolomonFridmanIJC15} to the stability and passivity analysis for diffusion partial differential equations with infinite distributed delays.
%\begin{equation}\label{Notations1:Inequality:infinite:doub:EF}
%\begin{array}{ll}
% K_0=\int_{0}^{+\infty} |K(s)| ds,\\
% K_1=\int_{0}^{+\infty} s|K(s)| ds,  \ \ hK_0+K_1 \bydef K_{1h}.
%\end{array}
%\end{equation}

To obtain more accurate lower bounds of integral inequalities (\ref{Inequality:infinite:EF}) and (\ref{Inequality:infinite:doub:EF}) over infinite intervals of integration, the method developed in \cite{SeuretAut13wirtinger} for the integral inequality over finite intervals of integration seems not to be applicable, since the function $f$ of (\ref{Inequality:Alex:f}) is directly dependent on both the lower limit $a$ and the upper limit $b$. Therefore, an interesting question arises:
\begin{question} Is it possible to derive more accurate lower bounds to reduce the conservativeness of integral inequalities (\ref{Inequality:infinite:EF}) and (\ref{Inequality:infinite:doub:EF})? If so, how much improvements can we obtain by applying the generalized inequalities to the stability analysis of continuous-time systems with gamma-distributed delays?
\end{question}
We further analyze the discrete-time case. Poisson-distribution is widespread in queuing theory \cite{BookGross08queueing}.
In \cite{Saleh1987}, the experimental data on the arrivals of pulses in indoor environments revealed that
each cluster's time-delay is poisson-distributed (see also \cite{Kotsopoulos08handbook}).
Therefore, we study the stability of linear discrete-time systems with poisson-distributed delays via appropriate Lyapunov functionals.
The Lyapunov-based analysis uses the discrete counterpart of integral inequalities (\ref{Inequality:infinite:EF}) and (\ref{Inequality:infinite:doub:EF}), i.e., Jensen inequalities with infinite sequences \cite{LiuYR08}, \cite{WuFuzzy12}.
%
%Our main challenge is the stability conditions for systems with gamma-distributed
%delays, where the delay is stabilizing, i.e., the corresponding system with the zero delay
%and the system without the delayed term are not asymptotically stable. Here
%the results are derived by using augmented Lyapunov functionals.
The following question corresponds to Question~1 in the discrete case:
\begin{question} Is it possible to generalize Jensen summation inequalities with infinite sequences?
If so, how much improvements can be achieved by applying the generalized inequalities to the stability analysis of discrete-time systems with poisson-distributed delays?
\end{question}
The central aim of the present paper is to answer the above questions.
%the first objective of the present paper is to derive a more general inequality, which
%corresponds to infinite intervals of integration and encompasses
%the inequality (\ref{Inequality:Alex}) that was introduced in \cite{SeuretAut13wirtinger}.
%In light of the above observation, by virtue of a more general auxiliary function $g(u)$  we first develop an integral inequality, which corresponds to finite intervals of integration and includes
%the one proposed in \cite{SeuretAut13wirtinger} as a particular case.
%Then by the same methodology
First, we present generalized Jensen integral inequality and its double integral extension, which are over infinite intervals of
integration. We show how they can be  applied to improve the stability result for linear continuous-time systems with gamma-distributed delays.
Then, for the discrete counterpart we provide extended Jensen summation inequality with infinite sequences, which leads to less conservative stability conditions for linear discrete-time systems with poisson-distributed delays.
In both the continuous-time and discrete-time cases,  the considered infinite distributed delays are shown to have stabilizing effects. Following \cite{SolomonFridmanAut13}, we derive the results via augmented Lyapunov functionals.

The structure of this paper is as follows.
%The current section will end with some notational conventions.
In Section \ref{sec:infinite:extend}
we derive generalized Jensen integral inequalities. %, which are on the infinite intervals of integration.
Section \ref{sec:sys:gamma} presents stability results for linear continuous-time systems with gamma-distributed delays to illustrate the efficiency of the
proposed inequalities.
Sections \ref{sec:infinite:extend:dis} and \ref{sec:sys:poisson} discuss the corresponding extended Jensen summation inequality with infinite sequences
and its application to the stability analysis of linear discrete-time systems with poisson-distributed delays, respectively.
%In Section \ref{sec:ex}, the efficiency of the presented approach is illustrated by an uncertain cart-pendulum
%example.
The conclusions and the future work will be stated in Section \ref{sec:conclusions}.

{\bf Notations:}\ The notations used throughout the paper are
standard. The superscript `$T$' stands
for matrix transposition, $ {\bbR}^n$ denotes the $n$ dimensional
Euclidean space with vector norm $|\cdot |$, $ {\bbR}^{n\times
m}$ is the set of all $n\times m$ real matrices, and the notation
$P \succ 0$, for  $P\in {\bbR}^{n\times n}$ means that  $P$ is
symmetric and positive  definite.
The symmetric term in a symmetric matrix is denoted by~${*}$.
The symbols ${\bbR},$ ${\bbR}^{+},$ ${\bbZ}^{+}$ and ${\bbN}$ denote the set of real numbers, non-negative real numbers, non-negative integers and
positive integers, respectively.

%\section{Improved inequalities with infinite intervals of integration}\label{sec:infinite:extend}
%\subsection{Extended Jensen inequality with finite intervals of integration}
%\section{Stability of systems with infinite delays via improved integral inequalities}
\section{Extended Jensen integral inequalities}\label{sec:infinite:extend}
The objective of this section is to provide extended Jensen integral inequalities over infinite intervals.
To do so, we first prove the generalized Jensen integral inequality introduced in \cite{ParkJFI2015} over finite intervals
in a simpler way. Then we extend the method to prove the inequality over infinite intervals.

\subsection{Extended Jensen integral inequality over finite intervals}
By changing $a+b-2u$ of (\ref{Inequality:Alex:f}) to a more
general scalar function $g(u)$ with
$\int_{a}^{b}g(u)du=0$ and $g(u)$ not identically zero, %equivalent to $0$ for all $u \in [a, b]$,
we %follow the proof
%of Theorem 3.6 in \cite{FridmanBook14} and
first present the extended Jensen inequality over finite intervals of integration. % as follows:
\begin{lemma} \cite{ParkJFI2015}  \label{lem:Inequality:Jensen:extend}
If there exist an  $n \times n$ matrix $R \succ 0,$ a scalar function
$g: [a, b] \to \bbR$ and a vector function
$\omega: [a, b] \to \bbR^n$
such that the integrations concerned are well
defined and $\int_{a}^{b}g(s) ds=0,$ where $g(s)$ is not identically zero, then the following inequality holds:
\begin{equation}
\label{Inequality:Jensen:extend}
\begin{array}{ll}
\int_{a}^{b} \omega^T(s) R \omega(s) ds
\geq {1 \over b-a} \int_{a}^{b}\omega^T(s) ds R
\int_{a}^{b} \omega(s) ds\\
\ \ \ +\Big[\int_{a}^{b} g^2(s) ds \Big]^{-1}
\int_{a}^{b}g(s)\omega^T(s) ds R \int_{a}^{b}g(s)\omega(s) ds.
\end{array}
\end{equation}
\end{lemma}
%%\hspace{16cm}$\blacksquare$\\
%%
%  %For any continuous function $\omega: [a, b] \to \bbR^n$ and any scalar function
%%$g: [a, b] \to \bbR$ with $\int_{a}^{b}g(s) ds=0$  and $g(s)$ not equivalent to $0$ for all $s \in [a, b],$ assume that the integrations concerned are well
%%defined.
{\bf Proof}: Define a function $f(u)$ for all
$u \in [a, b]$ by
\begin{equation}\label{Inequality:Jensen:extend:fs}
\begin{array}{ll}
f(u)=\omega(u)-{1 \over b-a} \int_{a}^{b}\omega(s) ds-g(u) \Theta,
\end{array}
\end{equation}
where $\Theta \in \bbR^n$ is a constant vector to be defined. Then, since $R \succ 0$ it follows that
\begin{equation}\nonumber %\label{Inequality:finite:proof:Jensen}
\begin{array}{llll}
 0 &\leq&  \int_a^b f^T(s) R f(s) ds\\
 &=&  \int_a^b \!\Big [\omega(s)\!-\!\!{1 \over b\!-\!a} \!\int_{a}^{b}\!\!\omega(\tet) d\tet\Big]^T R \Big [\omega(s)\!\!-\!\!{1 \over b\!-\!a} \! \int_{a}^{b}\!\!\omega(\tet) d\tet\Big]ds\\
 &&  + \Big [{2 \over b-a} \int_{a}^{b}\omega^T(s) ds R \Theta \Big] \int_{a}^{b}g(s) ds \\
&&  +\Big [\int_{a}^{b} g^2(s) ds \Big] \Theta^T R \Theta   -2 \Theta^T R \int_a^b g(s)\omega(s)ds.
\end{array}
\end{equation}
Noting that $\int_{a}^{b}g(s) ds=0$, we obtain
\begin{equation}\nonumber
\begin{array}{llll}
 \int_{a}^{b} \omega^T(s) R \omega(s) ds \geq
 {1 \over b-a} \int_{a}^{b}\omega^T(s) ds R
\int_{a}^{b} \omega(s) ds\\
\quad -\Big [\int_{a}^{b} g^2(s) ds \Big] \Theta^T R \Theta   +2 \Theta^T R \int_a^b g(s)\omega(s)ds.
\end{array}
\end{equation}
Rewriting the last two terms as sum of squares yields
\begin{equation}\label{Inequality:finite:proof:Jensen:2}
\begin{array}{llll}
 \int_{a}^{b} \omega^T(s) R \omega(s) ds \geq
 {1 \over b-a} \int_{a}^{b}\omega^T(s) ds R
\int_{a}^{b} \omega(s) ds\\
\quad  +\Big[\int_{a}^{b} g^2(s) ds \Big]^{-1}
\int_{a}^{b}g(s)\omega^T(s) ds R \int_{a}^{b}g(s)\omega(s) ds\\
\quad  -\Big [\int_{a}^{b} g^2(s) ds \Big] [\Theta-\Upsilon]^T R [\Theta-\Upsilon],
\end{array}
\end{equation}
where
\begin{equation} \nonumber
\Upsilon=\Big[\int_{a}^{b} g^2(s) ds\Big]^{-1}\int_a^b g(s)\omega(s) ds.
\end{equation}
Since (\ref{Inequality:finite:proof:Jensen:2}) holds independently of the choice of $\Theta$, we may choose $\Theta=\Upsilon$, which leads to
the maximum of the right-hand side of (\ref{Inequality:finite:proof:Jensen:2}) and thus, (\ref{Inequality:Jensen:extend}) holds. This concludes the proof.
\begin{remark}%\hspace{3cm}$\blacksquare$
In  \cite{ParkJFI2015}, the proof was more complicated as the corresponding construction of (\ref{Inequality:Jensen:extend:fs})
relied on an auxiliary function $\bar g(u),$ where $\bar g(u)$ satisfies $g(u)=\bar g(u)-{1 \over b-a} \int_{a}^{b}\bar g(s) ds.$

\end{remark}
%
%
%\begin{remark}
%Choose a scalar function $g(s)=c(a+b-2u), \ c \in \bbR \backslash \{0\},$ where $\int_{a}^{b}c(a+b-2u)du=0.$ Then we have
% %
%\begin{equation} \nonumber
%\begin{array}{llll}
% \int_{a}^{b} g^2(s) ds=\int_{a}^{b}  (a+b-2s)^2 c^2 ds={(b-a)^3 c^2 \over 3},
%\end{array}
%\end{equation}
%%
%and
%\begin{equation} \nonumber
%\begin{array}{llll}
%&& \int_{a}^{b}g(s)\omega(s) ds= \int_{a}^{b}c(a+b-2s) \omega(s) ds\\
%%&=& \int_{a}^{b}2c(b-s) \omega(s) ds-\int_{a}^{b}c(b-a) \omega(s) ds\\
%&=& 2c\int_{a}^{b}\int_{a}^s \omega(r) dr ds-\int_{a}^{b}c(b-a) \omega(s) ds\\
%&=& c(b-a)\Big [{2 \over b-a}\int_{a}^{b}\int_{a}^s \omega(r) dr ds-\int_{a}^{b}\omega(s) ds \Big].
%\end{array}
%\end{equation}
%As a consequence, the last term of the right-hand side of  (\ref{Inequality:Jensen:extend}) is reduced to
%\begin{equation} \nonumber
%\begin{array}{ll}
%\Big[\int_{a}^{b} g^2(s) ds \Big]^{-1}
%\int_{a}^{b}g(s)\omega^T(s) ds R \int_{a}^{b}g(s)\omega(s) ds\\
%={3 \over {b-a}} \Omega^T R \Omega,
%\end{array}
%\end{equation}
%which recovers the result of \cite{SeuretAut13wirtinger}. %, where $\Omega$ is given by (\ref{Omega:Alex}).
%%
%%the proposed inequality (\ref{Inequality:Jensen:extend}) includes (\ref{Inequality:Alex})
%%that introduced in \cite{SeuretAut13wirtinger} as a special case.
%\end{remark}
%

\subsection{Generalized Jensen integral inequalities over infinite intervals}
%\subsection{Generalized Jensen inequalities with infinite intervals of integration}
%This section will provide answer to Question 1.
%We extend the method for proving Lemma \ref{lem:Inequality:Jensen:extend} that has been considered on the finite intervals of integration, to the infinite ones:
We extend the method used for proving Lemma \ref{lem:Inequality:Jensen:extend} from finite intervals of integration to infinite ones in the following result.
\begin{theorem}
\label{thm:Inequality:infinite}
For a given $n \times n$ matrix $R \succ 0,$ scalar functions
$g: [0, +\infty) \to \bbR,$ $K: [0, +\infty) \to \bbR^{+}$ and a vector function
$\omega: [0, +\infty) \to \bbR^n,$
assume that the integrations concerned are well
defined and  $\int_{0}^{+\infty} K(s) g(s) ds = 0$ with  $g(s)$ not identically zero. Then the following inequality holds:
\begin{equation}
\label{Inequality:infinite}
\begin{array}{ll}
\int_{0}^{+\infty}K(s) \omega^T(s) R \omega(s) ds\\
\geq K_0^{-1} \int_{0}^{+\infty}K(s)  \omega^T(s) ds  R
\int_{0}^{+\infty}K(s)  \omega(s) ds\\
\hspace{.4cm}+\Big[\int_{0}^{+\infty} K(s) g^2(s)ds \Big]^{-1}
\bar \Omega^T R \bar \Omega,
\end{array}
\end{equation}
where
\begin{equation}\label{Notations1:Inequality:infinite}
\begin{array}{ll}
 K_0=\int_{0}^{+\infty} K(s) ds,\\
 \bar \Omega=\int_{0}^{+\infty} K(s) g(s)  \omega(s) ds.
\end{array}
\end{equation}
%and
%\begin{equation}\label{Notations2:Inequality:infinite}
%\begin{array}{ll}
% \int_{0}^{+\infty} |K(s)| g(s) ds \equiv 0.
%\end{array}
%\end{equation}
\end{theorem}
%{\bf Proof:}
%\hspace{16cm}$\blacksquare$\\
%
{\bf Proof}:
%To prove (\ref{Inequality:infinite}), %a function $f(s)$
%%has to be constructed such that $\int_{0}^{+\infty}|K(s)| \omega^T(s) R \omega(s) ds$
%%appears naturally in the future developments. Thus
%a necessary condition is that a function $\bar f(u)$ with a form of $\sqrt {K(u)} \omega(u)-{\tilde f}(u)$ has to be constructed, where ${\tilde f}(u): [0, +\infty) \to \bbR^n,$ is a continuous function
%to be defined.

%For any continuous function $\omega: [0, +\infty) \to \bbR^n$ and any scalar function
%$g: [0, +\infty) \to \bbR$, assume that the integrations concerned are well
%defined.
Define a function $\bar f(u)$ for all
$u \in [0, +\infty)$ by
\begin{equation}\nonumber
\begin{array}{ll}
\bar f(u)\!=\!\sqrt {K(u)} \Big [\omega(u)\!-\!K_0^{-1} \int_{0}^{+\infty}K(s)  \omega(s) ds\!-g(u) \bar \Theta \Big],
\end{array}
\end{equation}
where $\bar \Theta \in \bbR^n$ is a constant vector to be defined. Because  $R \succ 0$  we have
\begin{equation}\nonumber %\label{Inequality:infinite:proof:1}
\begin{array}{llll}
 0 \leq  \int_{0}^{+\infty} \bar f^T(s) R \bar f(s) ds\\
 = \! \int_{0}^{+\infty}\!\! \Big [\sqrt {K(s)} \omega(s)\!-\!\!K_0^{-1}\sqrt {K(s)}  \int_{0}^{+\infty}\!\!K(\tet)  \omega(\tet) d\tet\Big]^T \!R \\
  \hspace{.6cm}\times \Big [\sqrt {K(s)} \omega(s)\!\!-\!\!K_0^{-1}\sqrt {K(s)}  \int_{0}^{+\infty}K(\tet)  \omega(\tet) d\tet\Big]ds\\
  \hspace{.4cm}+ \Big[2 K_0^{-1} \bar \Theta^T R \int_{0}^{+\infty} K(s) \omega(s) ds \Big]  \int_{0}^{+\infty} K(s) g(s) ds \\
   \hspace{.4cm}+\Big [\int_{0}^{+\infty} K(s) g^2(s) ds\Big] \bar \Theta^T R \bar \Theta \\
  \hspace{.4cm} -2 \bar \Theta^T R \int_{0}^{+\infty} K(s) g(s) \omega(s)ds.
\end{array}
\end{equation}
Representing the last two terms as sum of squares together with $\int_{0}^{+\infty} K(s) g(s) ds=0$ yields
\begin{equation}\label{Inequality:infinite:proof:2}
\begin{array}{llll}
 \int_{0}^{+\infty}K(s) \omega^T(s) R \omega(s) ds\\
\geq K_0^{-1} \int_{0}^{+\infty}K(s)  \omega^T(s) ds  R
\int_{0}^{+\infty}K(s)  \omega(s) ds\\
 \hspace{.4cm}+\Big[\int_{0}^{+\infty} K(s) g^2(s)ds \Big]^{-1} \bar \Omega^T R \bar \Omega\\
 \hspace{.4cm} -\Big [\int_{0}^{+\infty} K(s) g^2(s) ds\Big] [\bar \Theta-\bar \Upsilon]^T R [\bar \Theta-\bar \Upsilon],
\end{array}
\end{equation}
where $\bar \Omega$ is given in (\ref{Notations1:Inequality:infinite}) and
\begin{equation} \nonumber
\bar \Upsilon =  \Big [\int_{0}^{+\infty} K(s) g^2(s) ds\Big]^{-1} \int_{0}^{+\infty}K(s)  g(s) \omega(s) ds.
\end{equation}
%
%
%Since (\ref{Inequality:infinite:proof:2}) holds independently of the choice of $\hat \Upsilon$,
Then, the same arguments in the proof of Lemma \ref{lem:Inequality:Jensen:extend}
and the choice $\bar \Theta=\bar \Upsilon$ lead to
the maximum of the right-hand side of (\ref{Inequality:infinite:proof:2}) and thus, (\ref{Inequality:infinite}) holds. This concludes the proof. %\hspace{3cm} $\blacksquare$
%%%%%%%%%%%%%%%%%%%%%%%%%%%%%%%%%%%%%%%%
%%%%%%%%%%%%%%%%%%%%%%%%%%%%%%%%%%%%%%%%

Note that the choice of $g(s)$ plays a crucial role in the application of Theorem \ref{thm:Inequality:infinite}.
Given $K_0$ in (\ref{Notations1:Inequality:infinite}) and
\begin{equation}\label{Inequality:infinite:K1}
\begin{array}{ll}
K_1=\int_{0}^{+\infty} sK(s) ds,
\end{array}
\end{equation}
let %employing (\ref{Notations2:Inequality:infinite})
\begin{equation}\label{Inequality:infinite:g}
\begin{array}{ll}
g(u)=c(K_0u-K_1), \ \ c \in \bbR \backslash \{0\}, \ \ u \geq 0,
\end{array}
\end{equation}
such that $\int_{0}^{+\infty} K(s) g(s) ds = 0$ holds.
Then, we find that
\begin{equation}\label{Inequality:infinite:g:2}
\begin{array}{ll}
 \int_{0}^{+\infty} K(s) g^2(s)ds=c^2\int_{0}^{+\infty} K(s) (K_0s-K_1)^2ds\\
\hspace{2.8cm} =c^2(K_0^2K_2-K_0K_1^2),\\
 \bar \Omega=\int_{0}^{+\infty} K(s) g(s)  \omega(s) ds\\
\ \ \ =c\Big[K_0\int_{0}^{+\infty} sK(s) \omega(s) ds-K_1\int_{0}^{+\infty} K(s) \omega(s) ds \Big],
\end{array}
\end{equation}
where
\begin{equation}\label{Inequality:infinite:K2}
\begin{array}{ll}
K_2=\int_{0}^{+\infty} s^2K(s) ds.
\end{array}
\end{equation}
%%%%%%%%%%%%%%%%%%%%%%%%%%%%%%%%%%%%%%%%%%%%
%%%%%%%%%%%%%%%%%%%%%%%%%%%%%%%%%%%%%%%%%%%%
From (\ref{Inequality:infinite:g}), (\ref{Inequality:infinite:g:2}) and Theorem \ref{thm:Inequality:infinite}, we have the following corollary:
%%%%%%%%%%%%%%%%%%%%%%%%%%%%%%%%%%%%%%
\begin{corollary}
\label{Cor:infinite}
For a given $n \times n$ matrix $R \succ 0,$ a scalar function
$K: [0, +\infty) \to \bbR^{+}$ and a vector function
$\omega: [0, +\infty) \to \bbR^n,$
assume that the integrations concerned are well
defined. Then the following inequality holds:
\begin{equation}
\label{Inequality:infinite:cor}
\begin{array}{ll}
\int_{0}^{+\infty}K(s) \omega^T(s) R \omega(s) ds\\
\hspace{.5cm}\geq K_0^{-1} \int_{0}^{+\infty}K(s)  \omega^T(s) ds  R
\int_{0}^{+\infty}K(s)  \omega(s) ds\\
\hspace{0.8cm}+\Big(K_2-{K_1^2 \over K_0} \Big)^{-1}
\tilde \Omega^T R \tilde \Omega,
\end{array}
\end{equation}
where $K_0,$  $K_1$ and $K_2$ are given by (\ref{Notations1:Inequality:infinite}), (\ref{Inequality:infinite:K1}) and (\ref{Inequality:infinite:K2}), respectively, and
\begin{equation} \nonumber
\begin{array}{ll}
%
%K_2=\int_{0}^{+\infty} s^2|K(s)| ds,\\
\tilde \Omega={K_1 \over K_0}\int_{0}^{+\infty} K(s) \omega(s) ds-\int_{0}^{+\infty} sK(s) \omega(s) ds.
\end{array}
\end{equation}
\end{corollary}
%\hspace{16cm}$\blacksquare$
%
%\section{Generalized double integral inequality with infinite intervals of integration}
%\subsubsection{Generalized double integral inequality with infinite intervals of integration}
The same methodology to prove Lemma \ref{lem:Inequality:Jensen:extend} and Theorem~\ref{thm:Inequality:infinite} can be applied to
generalize the inequality (\ref{Inequality:infinite:doub:EF}). We have the following result: % (see Appendix A for the proof):
\begin{theorem}
\label{thm:Inequality:infinite:doub}
If there exist an $n \times n$ matrix $R \succ 0,$ scalar functions
$g: [t-\tet-h, t] \to \bbR,$
$K: [0, +\infty) \to \bbR^{+},$ a scalar $h \geq 0$ and a vector function
$\omega: [t-\tet-h, t] \to \bbR^n$
such that the integrations concerned are well
defined and $\int_{0}^{+\infty} \int_{t-\tet-h}^{t}K(\tet) g(s) ds d \tet =0,$ where $g(s)$ is not identically zero, then the following inequality holds:
\begin{equation}
\label{Inequality:infinite:doub}
\begin{array}{ll}
\int_{0}^{+\infty} \int_{t-\tet-h}^{t}K(\tet) \omega^T(s) R \omega(s) ds d \tet \\
\hspace{.5cm}\geq K_{1h}^{-1} \int_{0}^{+\infty} \int_{t-\tet-h}^{t} K(\tet)  \omega^T(s) ds d \tet R \\
\hspace{1cm}\times \int_{0}^{+\infty}\int_{t-\tet-h}^{t} K(\tet)   \omega(s) ds  d \tet\\
\hspace{0.8cm}+\Big[\int_{0}^{+\infty}\int_{t-\tet-h}^{t} K(\tet) g^2(s)ds d \tet \Big]^{-1}
\Sigma^T R \Sigma,
\end{array}
\end{equation}
where %$K_{1h}=\int_{0}^{+\infty} |K(s)| (s+h)ds=hK_0+K_1$ and
\begin{equation}\label{Notations1:Inequality:infinite:doub}
\begin{array}{ll}
% K_0=\int_{0}^{+\infty} |K(s)| ds,\\
K_{1h}=\int_{0}^{+\infty} K(s) (s+h)ds=hK_0+K_1,\\
 \Sigma=\int_{0}^{+\infty}\int_{t-\tet-h}^{t} K(\tet) g(s)  \omega(s) ds d \tet.
\end{array}
\end{equation}
%and
%\begin{equation}\label{Notations2:Inequality:infinite:doub}
%\begin{array}{ll}
% \int_{0}^{+\infty} \int_{t-\tet-h}^{t}|K(\tet)| g(s) ds d \tet \equiv 0.
%\end{array}
%\end{equation}
\end{theorem}
{\bf Proof}: See Appendix A.
\begin{remark}
Theorems \ref{thm:Inequality:infinite}  and \ref{thm:Inequality:infinite:doub} refine the inequalities of \cite{SolomonFridmanAut13}, in which the last terms of the right-hand-side of  (\ref{Inequality:infinite})  and (\ref{Inequality:infinite:doub}) are zero.
Hence, our new inequalities develop more accurate lower bounds of $\int_{0}^{+\infty}K(s) \omega^T(s) R \omega(s) ds$ and $\int_{0}^{+\infty} \int_{t-\tet-h}^{t}K(\tet) \omega^T(s) R \omega(s) ds d \tet$ than the ones provided in \cite{SolomonFridmanAut13}.
\end{remark}
%%%%%%%%%%%%%%%%%%%%%%%%%%%%%%%%%%%%%%%%
%%%%%%%%%%%%%%%%%%%%%%%%%%%%%%%%%%%%%%%%
We choose a scalar function
%$g$ %employing (\ref{Notations2:Inequality:infinite:doub})
\begin{equation}\label{Inequality:infinite:doub:g}
\begin{array}{ll}
g(u)=-u+t-{h K_{1h}+hK_1+K_2 \over 2 K_{1h}},
\end{array}
\end{equation}
such that $\int_{0}^{+\infty} \int_{t-\tet-h}^{t}K(\tet) g(s) ds d \tet=0$,
where $K_1,$ $K_2$ and $K_{1h}$ are given by (\ref{Inequality:infinite:K1}), (\ref{Inequality:infinite:K2}) and (\ref{Notations1:Inequality:infinite:doub}), respectively.
%\begin{equation}
%\begin{array}{ll}
%
%K_2=\int_{0}^{+\infty} s^2|K(s)| ds.
%\end{array}
%\end{equation}
Hence, we have
\begin{equation}\label{Inequality:infinite:doub:g:2}
\begin{array}{llll}
\int_{0}^{+\infty}\int_{t-\tet-h}^{t} K(\tet) g^2(s)ds d \tet\\
=\int_{0}^{+\infty}\int_{t-\tet-h}^{t} K(\tet) \Big(-s+t-{h K_{1h}+hK_1+K_2 \over 2 K_{1h}}\Big)^2dsd \tet\\
=  {h^3 \over 2}K_0+2K_3+{3h^2K_0(hK_1+2K_2)-3K_2^2\over 2 K_{1h}} \bydef {\tilde K}_1,
\end{array}
\end{equation}
and
\begin{equation}\label{Inequality:infinite:doub:Sigma}
\begin{array}{llll}
\Sigma=  \int_{0}^{+\infty}\int_{t-\tet-h}^{t} K(\tet) g(s)  \omega(s) ds d \tet\\
=  \int_{0}^{+\infty}\int_{t-\tet-h}^{t}\! K(\tet) \Big(-\!s\!+\!t\!-\!{h K_{1h}+hK_1+K_2 \over 2 K_{1h}}\Big)  \omega(s) ds d \tet\\
= \int_{0}^{+\infty}\!\!\int_{t-\tet-h}^{t} \int_{s}^{t}K(\tet)   \omega(s) dr ds d \tet\!\\
\hspace{.5cm}- \Big({h \over 2}\!+\!{hK_1\!+\!K_2 \over 2 K_{1h}}\Big)\int_{0}^{+\infty}\!\int_{t-\tet-h}^{t} \!K(\tet)   \omega(s) ds  d \tet\\
= \int_{0}^{+\infty}\!\int_{t-\tet-h}^{t} \int_{t-\tet-h}^{r}\!K(\tet)   \omega(s) ds dr d \tet\!\\
\hspace{.5cm}- \Big({h \over 2}\!+\!{hK_1\!+\!K_2 \over 2 K_{1h}}\Big)\int_{0}^{+\infty}\!\int_{t-\tet-h}^{t} \!K(\tet)   \omega(s) ds  d \tet \bydef \tilde \Sigma,
\end{array}
\end{equation}
where
\begin{equation}\nonumber %\label{Inequality:infinite:doub:K3}
\begin{array}{ll}
K_3=\int_{0}^{+\infty} s^3K(s) ds.
\end{array}
\end{equation}
From (\ref{Inequality:infinite:doub:g})--(\ref{Inequality:infinite:doub:Sigma}) and Theorem \ref{thm:Inequality:infinite:doub}, we arrive at the following result:
%%%%%%%%%%%%%%%%%%%%%%%%%%%%%%%%%%%%%%
\begin{corollary}
\label{Cor:infinite:doub}
If there exist an $n \times n$ matrix $R \succ 0,$ a scalar function
$K: [0, +\infty) \to \bbR^{+},$ a scalar $h \geq 0$ and a vector function
$\omega: [t-\tet-h, t] \to \bbR^n$
such that the integrations concerned are well
defined, then the following inequality holds:
\begin{equation}
\label{Inequality:infinite:cor:doub}
\begin{array}{ll}
\int_{0}^{+\infty} \int_{t-\tet-h}^{t}K(\tet) \omega^T(s) R \omega(s) ds d \tet \\
\hspace{1cm}\geq K_{1h}^{-1} \int_{0}^{+\infty} \int_{t-\tet-h}^{t} K(\tet)  \omega^T(s) ds d \tet R\\
\hspace{1.5cm}\times \int_{0}^{+\infty}\int_{t-\tet-h}^{t} K(\tet)   \omega(s) ds  d \tet\\
\hspace{1.3cm}+{\tilde K}_1^{-1}
\tilde \Sigma^T R \tilde \Sigma,
\end{array}
\end{equation}
where ${\tilde K}_1$ and $\tilde \Sigma$ are given by (\ref{Inequality:infinite:doub:g:2}) and (\ref{Inequality:infinite:doub:Sigma}), respectively.
%\begin{equation}
%\begin{array}{ll}
%
%K_2=\int_{0}^{+\infty} s^2|K(s)| ds,\ \ \
%
%K_3=\int_{0}^{+\infty} s^3|K(s)| ds.
%%
%%{\tilde K}_1={h^3 \over 2}K_0+2K_3+{3h^2(hK_1+2K_2)-3K_2^2\over 2 K_{1h}}.
%%
%%\tilde \Sigma=\Big({h \over 2}+{hK_1+K_2 \over 2 K_{1h}}\Big)\int_{0}^{+\infty}\int_{t-\tet-h}^{t} K(\tet)   \omega(s) ds  d \tet-\int_{0}^{+\infty}\int_{t-\tet-h}^{t} \int_{t-\tet-h}^{r}K(\tet)   \omega(s) ds dr d \tet.
%\end{array}
%\end{equation}
\end{corollary}
The generalized integral inequality (\ref{Inequality:infinite:cor}) and its double integral extension (\ref{Inequality:infinite:cor:doub}) will be employed for the stability analysis of continuous-time systems with gamma-distributed delays in the next section.
%\hspace{16cm}$\blacksquare$
\section{Stability analysis of continuous-time systems with gamma-distributed delays} \label{sec:sys:gamma}
%%%%%%%%%%%%%%%%%%%%%%%%%%%%
%In this section, the space of continuous functions $\phi:$ $(-\infty, 0] \to {\bbR}^{n}$ with the norm $\| \phi\|_C=sup_{s \in (-\infty, 0]}|\phi(s)| <+\infty$
%is denoted by $C(-\infty, 0]$. The space of the continuously differentiable functions $\phi:$ $(-\infty, 0] \to {\bbR}^{n}$ with the norm $\| \phi\|_{C^1}=\| \phi\|_C +\| \dot \phi\|_C <+\infty$
%is denoted by $C^1(-\infty, 0]$.
%
%%%%%%%%%%%%%%%%%%%%%%%%%%%%%%%%%%%%%%%%
Consider the linear continuous-time systems with gamma-distributed delays:
\begin{equation}
\label{sys:gamma}
\begin{array}{lll}
\dot x(t)=Ax(t)+A_1 \int_{0}^{+\infty} \Gamma(\tet) x(t-\tet-h) d\tet,
\end{array}
\end{equation}
where $x(t)\in  {\bbR}^{n}$ is the state vector, $A,$ $A_1 \in  {\bbR}^{n \times n}$  are constant system matrices, and
$h \geq 0$ represents a fixed time gap. The smooth kernel $\Gamma$ is given by $\Gamma(\tet)={\tet^{N-1}e^{-{\tet \over T}}\over T^N (N-1)!},$
where $N \geq 2,$ $N \in \bbN,$ is a shape parameter of the distribution and $T>0$ is a scale parameter.
The matrices $A$ and $A+A_1$ are not allowed to be Hurwitz. The initial condition is given by $\phi \in C^1(-\infty, 0],$ where
$C^1(-\infty, 0]$ denotes the space of continuously differentiable functions $\phi:$ $(-\infty, 0] \to {\bbR}^{n}$ with the norm $\| \phi\|_{C^1}=\| \phi\|_C +\| \dot \phi\|_C <+\infty, $ $\| \phi\|_C=\sup_{s \in (-\infty, 0]}|\phi(s)| <+\infty$.

Following \cite{SolomonFridmanAut13} and introducing
\begin{equation}
\begin{array}{lll} \nonumber
y(t)\!=\!\int_{0}^{+\infty} \Gamma(\tet) x(t-\!\tet\!-h) d\tet\!=\!\int_{-\infty}^{t} \Gamma(t\!-s) x(s\!-h) ds,
\end{array}
\end{equation}
the system (\ref{sys:gamma}) can be transformed into
\begin{equation}
\label{clc:aug}
\begin{array}{lll}
\dot x(t)&=&Ax(t)+A_1y(t),\\
\dot y(t)&=& -{1 \over T}y(t)+\rho(t),
\end{array}
\end{equation}
where
\begin{equation} \nonumber
\begin{array}{lll}
\rho(t)=\int_{-\infty}^{t} \Psi(t\!-s) x(s\!-h) ds=\int_{0}^{+\infty} \Psi(\tet) x(t-\!\tet\!-h) d\tet,\\
 \Psi(\tet)={\tet^{N-2}e^{-{\tet \over T}}\over T^N (N-2)!}.
\end{array}
\end{equation}
It follows readily that
\begin{equation} \label{Psi}
\begin{array}{lll}
\int_{0}^{+\infty} \Psi(\tet) d\tet ={1 \over T}\bydef \Psi_0,\\
 \int_{0}^{+\infty} \tet \Psi(\tet) d\tet =N-1\bydef \Psi_1,\\
 \int_{0}^{+\infty} \tet^2\Psi(\tet) d\tet=N(N-1)T \bydef \Psi_2,\\
 \int_{0}^{+\infty} \tet^3\Psi(\tet) d\tet=(N+1)N(N-1)T^2 \bydef \Psi_3,\\
 \int_{0}^{+\infty} (\tet+h)\Psi(\tet) d\tet=h\Psi_0+\Psi_1=N-1+{h \over T} \bydef \Psi_{1h}.\\
\end{array}
\end{equation}
In the following, we provide two sufficient conditions for the stability of system (\ref{clc:aug});
one is derived by applying (\ref{Inequality:infinite:cor}) and (\ref{Inequality:infinite:doub:EF}), the other is obtained from (\ref{Inequality:infinite:cor}) and (\ref{Inequality:infinite:cor:doub}).
%An example below for the traffic flow model on the ring illustrates
%the efficiency of the given conditions.

\subsection{Stability result I}
We consider the following augmented LKF:
\begin{equation}\label{V}
\begin{array}{lll}
V(t)=V_1(t)+V_{G}(t)+V_{H}(t),\\
V_1(t)=\eta^T(t) W \eta (t),\\
V_{G}(t)= \int_{0}^{+\infty} \int_{t-\tet-h}^{t} \Psi(\tet) x^T(s) G x(s)ds d \tet,\\
V_{H}(t)= \int_{0}^{+\infty} \int_{0}^{\tet+h} \int_{t-\lambda}^{t} \Psi(\tet) {\dot x}^T(s) H {\dot x}(s)ds d \lambda d \tet,\\
\end{array}
\end{equation}
where $W \succ 0, G \succ 0, \ H \succ 0, $ $\eta(t)=\col\{x(t), y(t)\}$.
%
%\begin{equation}\label{W}
%\begin{array}{lll}
%W=\left[\begin{array} {ccc}
%P & Q\\
%{*} & Z
%\end{array} \right] \succ 0.
%\end{array}
%\end{equation}
%
Since $A$ and $A+A_1$ are not allowed to be Hurwitz,  we use augmented Lyapunov functionals.
The term $V_H$ extends the triple integral of \cite{SunJianIJRNC09} for finite delay to infinite delay \cite{SolomonFridmanAut13}.
\begin{remark}
The recent method of \cite{SolomonFridmanAut13} for the stability analysis of system (\ref{clc:aug}) is based on a functional of the form
\begin{equation} \label{V:SolomonFridmanAut13}
\begin{array}{lll}
V(t)=V_1(t)+V_{G}(t)+V_{H}(t)+V_{E}(t)+V_{F}(t),\\
V_{E}(t)= \int_{0}^{+\infty} \int_{t-\tet-h}^{t} \Gamma(\tet) x^T(s) E x(s)ds d \tet,\\
V_{F}(t)= \int_{0}^{+\infty} \int_{0}^{\tet+h} \int_{t-\lambda}^{t} \Gamma(\tet) {\dot x}^T(s) F {\dot x}(s)ds d \lambda d \tet
\end{array}
\end{equation}
together with the utilization of the integral inequalities (\ref{Inequality:infinite:EF}) and (\ref{Inequality:infinite:doub:EF}).
Compared to (\ref{V}), the functional (\ref{V:SolomonFridmanAut13}) has two additional terms $V_{E}(t)$ and $V_{F}(t)$.
%The Lyapunov-based analysis was based on integral inequalities (\ref{Inequality:infinite:EF}) and (\ref{Inequality:infinite:doub:EF}).
In the example below, we will show the advantages of our proposed approach (larger stability region in the
$(T, h) $ plane and less number of scalar decision variables).
The improvement is achieved due to that the application of Corollary \ref{Cor:infinite} leads to one more negative term in the derivative of the LKF.
\end{remark}

The following proposition is provided for the asymptotic stability of system (\ref{clc:aug}).
\begin{proposition}
\label{thm:gamma:N:2}
If there exist  $2n\times 2n$ positive definite matrix $W$ and $n\times n$ positive definite matrices $G,H$ such that the following LMI
is feasible:
\begin{equation} \label{Xi:thm:gamma:extend}
\begin{array}{lll}
\Xi=\Sigma+F_1^T W F_0+F_0^T W F_1-{N-1 \over T}F_{23}^T G F_{23} \\
\hspace{.7cm}+ \Psi_{1h} F_{01}^T H F_{01}-\Psi_{1h}^{-1}
F_{13}^T H F_{13} \prec 0,
\end{array}
\end{equation}
where $\Psi_{1h}$ is given by (\ref{Psi}), $\Sigma={\rm diag}\{{1 \over T} G, 0 , -TG\}$ and
\begin{equation} \nonumber %\label{Notations:thm:gamma:extend:single}
\begin{array}{lll}
%\Sigma={\rm diag}\{{1 \over T} G, 0 , TG\},\\
F_0=\left[\begin{array} {ccccc}
A & A_1 & 0 \\
0 & -{1 \over T} & I
\end{array} \right],\ \
F_1=\left[\begin{array} {ccccc}
I & 0 & 0 \\
0 & I & 0
\end{array} \right],\\
F_{01}=[A \ A_1 \ 0],\
F_{13}=[{1 \over T}I \ 0 \ -I],\
F_{23}=[0 \ -I \ \ TI],
\end{array}
\end{equation}
then system (\ref{clc:aug}) is asymptotically stable. %\hspace{9cm}$\blacksquare$
\end{proposition}
{\bf Proof:}
Differentiating $V_1(t)$ along (\ref{clc:aug}), we have
\begin{equation}
\begin{array}{lll} \nonumber
\dot V_1(t)=2\eta^T(t) W \dot {\eta}(t)=2 \xi^T(t)F_1^T W F_0  \xi(t)
\end{array}
\end{equation}
with $\xi(t)=\col \{ x(t), \ y(t), \ \rho(t)\}$.
By applying Corollary~\ref{Cor:infinite} we obtain
\begin{equation}
\begin{array}{lll} \label{dotVG}
{\dot V_G}(t)\!=\! \Psi_0 x^T(t) G x(t)\!\!-\!\! \int_{0}^{+\infty} \!\!\Psi(\tet) x^T(t\!-\!\tet\!-\!\!h)G x(t\!-\!\tet\!\!-\!\!h) d\tet\\
% &\leq& \Psi_0 x^T(t) G x(t)- \Psi_0^{-1}\rho^T(t)G \rho(t)-\Big(\Psi_2-{\Psi_1^2 \over \Psi_0} \Big)^{-1} \times\\
%&&\Big[{\Psi_1 \over \Psi_0} \int_{0}^{+\infty} \Psi(\tet) x(t-\tet-h) d\tet- \int_{0}^{+\infty} \tet \Psi(\tet) x(t-\tet-h) d\tet ds\Big ]^TG\\
%&&\Big[{\Psi_1 \over \Psi_0} \int_{0}^{+\infty} \Psi(\tet) x(t-\tet-h) d\tet- \int_{0}^{+\infty} \tet \Psi(\tet) x(t-\tet-h) d\tet ds\Big ]\\
\hspace{.9cm} \leq \Psi_0 x^T(t) G x(t)- \Psi_0^{-1}\rho^T(t)G \rho(t)\\
\hspace{1.5cm}-\Big(\Psi_2-{\Psi_1^2 \over \Psi_0} \Big)^{-1}
\Big[{\Psi_1 \over \Psi_0}\rho(t)-(N-1)y(t)\Big]^T G\\
\hspace{1.8cm} \times \Big[{\Psi_1 \over  \Psi_0}\rho(t)-(N-1)y(t)\Big]\\
\hspace{.9cm}= {1 \over T} x^T(t) G x(t)- T\rho^T(t)G \rho(t)\\
\hspace{1.4cm} -{N-1 \over T}
\Big[T\rho(t)-y(t)\Big]^T G \Big[T\rho(t)-y(t)\Big]\\
\hspace{.9cm}= {1 \over T} x^T(t) G x(t)- T\rho^T(t)G \rho(t)\\
\hspace{1.4cm} -{N-1 \over T}\xi^T(t)F_{23}^T G F_{23}  \xi(t).
\end{array}
\end{equation}
Furthermore, applying (\ref{Inequality:infinite:doub:EF}) we find that
\begin{equation} \nonumber
\begin{array}{lll}
{\dot V_H}(t)&\leq & %(h\Psi_0+\Psi_1) {\dot x}^T(t) H {\dot x}(t)-\Big(h\Psi_0+\Psi_1 \Big)^{-1}
%\Big[ \Psi_0 x(t)-\rho(t)\Big]^T H \Big[\Psi_0 x(t)-\rho(t)\Big],\\
 \Psi_{1h} {\dot x}^T(t) H {\dot x}(t)\\
 &&-\Psi_{1h}^{-1}
\Big[ \Psi_0 x(t)-\rho(t)\Big]^T H \Big[\Psi_0 x(t)-\rho(t)\Big]\\
&=& \xi^T(t) [\Psi_{1h} F_{01}^T H F_{01}-\Psi_{1h}^{-1}
F_{13}^T H F_{13}]  \xi(t).
\end{array}
\end{equation}
Therefore, (\ref{Xi:thm:gamma:extend}) guarantees that $\dot V(t) \leq \xi^T(t) \Xi  \xi(t) \leq - \beta | x(t)|^2$ for some $\beta>0$, which
proves the asymptotic stability. %\hspace{10.8cm}$\blacksquare$
\subsection{Stability result II} $\newline$
%In this case, we suggest the LKF given by: %the term $V_1$ in (\ref{V}) is modified as follows:
%
The stability of system (\ref{clc:aug}) can be alternatively analyzed via a LKF given by
\begin{equation}\nonumber %\label{V:2}
\begin{array}{lll}
\bar V(t)=\bar V_1(t)+V_{G}(t)+V_{H}(t),\ \ \bar V_1(t)=\bar \eta^T(t) \bar W \bar \eta(t),
\end{array}
\end{equation}
where $\bar W \succ 0,$  $\bar \eta(t)=\col\{x(t), \ y(t), \ \zeta(t)\},$  $\zeta(t)= \int_{0}^{+\infty} \int_{t-\tet-h}^{t}\Psi(\tet) x(s)ds d \tet,$ $V_{G}(t)$ and $V_{H}(t)$ are given by (\ref{V}).
%%
%\begin{equation}\label{W:2}
%\begin{array}{lll}
%\bar W=\left[\begin{array} {ccc}
%P & Q_1 & Q_2\\
%{*} & Z & Q_3\\
%{*} & {*} & R
%\end{array} \right] \succ 0.
%\end{array}
%\end{equation}
%
%
Noting that $\dot \zeta(t)={1 \over T} x(t)-\rho(t)$
% with $\Psi_0$ given by (\ref{Psi}).
and differentiating $V_1(t)$ along (\ref{clc:aug}), we have
\begin{equation}
\begin{array}{lll} \label{dotV1:2}
\dot {\bar V}_1(t)&=&2 \bar \eta^T(t) \bar W \left[\begin{array} {ccc}
Ax(t)+A_1y(t)\\
-{1 \over T}y(t)+\rho(t)\\
{1 \over T} x(t)-\rho(t)
\end{array} \right]\\
&=& 2 \bar  \xi^T(t) \bar F_1^T \bar W \bar F_0 \bar  \xi(t)
\end{array}
\end{equation}
with   $\bar \xi(t)=\col \{ x(t), \ y(t), \ \rho(t), \ \zeta(t)\}$ and
\begin{equation}\label{barF1}
\begin{array}{lll}
\bar F_1=\left[\begin{array} {ccccc}
I & 0 & 0 & 0\\
0 &I  & 0 & 0\\
0 &0  & 0 & I
\end{array} \right],\
\bar F_0=\left[\begin{array} {ccccc}
A & A_1 & 0 & 0\\
0 &-{1 \over T}I  & I & 0\\
{1 \over T}I &0  & -I & 0
\end{array} \right].
\end{array}
\end{equation}
Furthermore, by applying (\ref{Inequality:infinite:cor:doub}) we find that
\begin{equation}
\begin{array}{lll} \label{dotVH:2}
{\dot V_H}(t)%&\leq & (h\Psi_0+\Psi_1) {\dot x}^T(t) H {\dot x}(t)-\Big(h\Psi_0+\Psi_1 \Big)^{-1}
%\Big[ \Psi_0 x(t)-\rho(t)\Big]^T H \Big[\Psi_0 x(t)-\rho(t)\Big]\\
%&& -{\tilde \Psi}_1^{-1}\varphi^T(t) H \varphi(t)\\
\leq \Psi_{1h} {\dot x}^T(t) H {\dot x}(t)\\
\hspace{1.3cm}-\Psi_{1h}^{-1}
\Big[ \Psi_0 x(t)-\rho(t)\Big]^T H \Big[\Psi_0 x(t)-\rho(t)\Big]\\
\hspace{1.3cm}-{\tilde \Psi}_1^{-1}\varphi^T(t) H \varphi(t)\\
\hspace{.1cm}= \bar \xi^T(t) [\Psi_{1h} \bar F_{01}^T H \bar F_{01}\!\!-\!\!\Psi_{1h}^{-1}
\bar F_{13}^T H \bar F_{13}\!\!-\!\!{\tilde \Psi}_1^{-1}\bar F_{33}^T H \bar F_{33}] \bar \xi(t),
\end{array}
\end{equation}
where $\Psi_{1h}$ is given in (\ref{Psi}) and
\begin{equation}\label{barF01}
\begin{array}{rlllll}
\bar F_{01}&=&[A \ A_1 \ 0 \ 0],\
\bar F_{13}=[{1 \over T}I \ 0 \ -I \ 0],\\
\varphi(t)&=&\hbar [\Psi_0 x(t)-\rho(t)]-\zeta(t)+h \rho(t)+(N-1)y(t)\\
&=&\bar F_{33}\bar \xi(t),\\
\bar F_{33}&=&[{\hbar \over T}I \ (N-1)I \ (h-\hbar)I \ -I],\\
{\tilde \Psi}_1&=&{h^3 \over 2T}+2\Psi_3+{3h^2\Psi_0(h\Psi_1+2\Psi_2)-3\Psi_2^2\over 2 \Psi_{1h}},\\
\hbar&=&\Big({h \over 2}+{h\Psi_1+\Psi_2 \over 2 \Psi_{1h}}\Big).
\end{array}
\end{equation}
Therefore, by combining (\ref{dotVG}), (\ref{dotV1:2}) and (\ref{dotVH:2}) we obtain $\dot {\bar V}(t) \leq \bar \xi^T(t) \bar \Xi \bar \xi(t) \leq - \bar \beta | x(t)|^2$ for some $\bar \beta>0$, if
\begin{equation} \label{barXi:thm:gamma:extend}
\begin{array}{lll}
\bar \Xi=\bar \Sigma+\bar F_1^T \bar W \bar F_0+\bar F_0^T \bar W \bar F_1-{N-1 \over T}\bar F_{23}^T G \bar F_{23} \\
\quad \ + \Psi_{1h} \bar F_{01}^T H \bar F_{01}
 -\Psi_{1h}^{-1}
\bar F_{13}^T H \bar F_{13}-{\tilde \Psi}_1^{-1}\bar F_{33}^T H \bar F_{33} \prec 0,
\end{array}
\end{equation}
where
\begin{equation} \label{barF23}
\begin{array}{lll}
\bar \Sigma&=&{\rm diag}\{{1 \over T} G, 0 , -TG, 0\},\\
\bar F_{23}&=&[0 \ -I \ \ TI \ 0].
\end{array}
\end{equation}
%
%\begin{equation}\label{Xi:2}
%\begin{array}{lll}
%\bar \Xi=\left[\begin{array} {ccccccc}
%\bar \Xi_{11} & \bar \Xi_{12} & \bar \Xi_{13} & A^TQ_2+\Psi_0 R+{\tilde \Psi}_1^{-1}\Psi_0 \hbar H & \Psi_{1h}A^TH\\
%{*} &\bar \Xi_{22} & \bar \Xi_{23} & A_1^TQ_2-{1 \over T}Q_3+{\tilde \Psi}_1^{-1}(N-1) H &\Psi_{1h}A_1^TH\\
%{*} &{*}  & \bar \Xi_{33} & Q_3-R+{\tilde \Psi}_1^{-1} (h-\hbar)H& 0\\
%{*} &{*}  & {*} & -{\tilde \Psi}_1^{-1}H  & 0\\
%{*} &{*}  & {*} & {*} &  -\Psi_{1h}H
%\end{array} \right]<0,
%\end{array}
%\end{equation}
%where
%\begin{equation}\label{Xi:notations:2}
%\begin{array}{lll}
%\bar \Xi_{11}&=&PA+A^TP+\Psi_0 Q_2+\Psi_0 Q_2^T+{1 \over T}G-{\Psi_0^2 \over \Psi_{1h}}H-{\tilde \Psi}_1^{-1}\Psi_0^2 \hbar^2H,\\
%\bar \Xi_{12}&=& PA_1-{1 \over T}Q_1+A^TQ_1+\Psi_0 Q_3^T-{\tilde \Psi}_1^{-1}\Psi_0 \hbar (N-1)H,\\
%\bar \Xi_{13}&=& Q_1-Q_2+{\Psi_0 \over \Psi_{1h}}H-{\tilde \Psi}_1^{-1}\Psi_0 \hbar (h-\hbar)H,\\
%\bar \Xi_{22}&=&
%Q_1^TA_1+A_1^TQ_1-{2 \over T}Z- {N-1 \over T}G-{\tilde \Psi}_1^{-1}(N-1)^2H,\\
%\bar \Xi_{23}&=& Z-Q_3+(N-1)G-{\tilde \Psi}_1^{-1}(N-1)(h-\hbar)H ,\\
%\bar \Xi_{33}&=&-NTG-\Psi_{1h}^{-1}H-{\tilde \Psi}_1^{-1} (h-\hbar)^2H.
%%\Xi_{23}&=&Z+\Big(\Psi_2-{\Psi_1^2 \over \Psi_0} \Big)^{-1}{ \Psi_1 \over  \Psi_0}(N-1)G=Z+(N-1)G,\\
%%\Xi_{33}&=&- \Psi_0^{-1}G-\Big(\Psi_2-{ \Psi_1^2 \over  \Psi_0} \Big)^{-1}{ \Psi_1^2 \over  \Psi_0^2}G-\Big(h\Psi_0+\Psi_1 \Big)^{-1}H\\
%%%&=&-TG-(N-1)TG-\Psi_{1h}^{-1}H\\
%%&=&-NTG-\Psi_{1h}^{-1}H.
%\end{array}
%\end{equation}
We have thus proved the following proposition:
\begin{proposition}
\label{thm2:gamma:N:2}
If there exist $3n\times 3n$ positive definite matrix $\bar W$ and $n\times n$ positive definite matrices $G,H$ such that LMI (\ref{barXi:thm:gamma:extend}) with notations given by
(\ref{Psi}), (\ref{barF1}), (\ref{barF01}) and (\ref{barF23}) is feasible,
then system (\ref{clc:aug}) is asymptotically stable. %\hspace{7cm}$\blacksquare$
\end{proposition}

Next we present an example to illustrate the applicability of the theoretical results.

\subsection{Example 1}
%************************ m-files:
%*****************************
% 18-gamma//Gamma_N_2_single_Compact.m %%%%%%%%%%  Kun, use extended single
% 18-gamma//Gamma_N_2_extended_double_Compact.m %%%%%%%%%%  Kun, use extended double
% 18-gamma//Gamma_N_2_emiliaAut_Prop3.m %%%%%%%%%%  EmiliaAut13, Prop.3
%Plot!!!*****************************
% 18-gamma//Plot_gamma_kun_New %%%%%%%%%%  Main.m
We illustrate the efficiency of the presented results through an example of two cars on a
ring, see \cite{MorarescuSIAM07} and \cite{SolomonFridmanAut13} for details. In this example,
\begin{equation} \nonumber
A=0 \ \  \text{and} \ \
A_1= \left[
\begin{array}{cc}
       -2& 2 \\
       2 & -2
      \end{array}\right],
\end{equation}
so neither $A$ nor $A+A_1$ is Hurwitz.
For the values of $h$ given in Table I and $N=2$, by applying the method in \cite{SolomonFridmanAut13} and
using Propositions \ref{thm:gamma:N:2} and \ref{thm2:gamma:N:2}, we obtain the maximum allowable values of $T$ that achieve the stability.
Fig. 1 presents tradeoff curves between maximal allowable $T$ and $h$ by applying the above three methods.
Furthermore, the stability region in the
$(T, h)$ plane that preserves the asymptotic stability is depicted in Figs.~2--4 by using the condition in \cite{SolomonFridmanAut13}, Proposition \ref{thm:gamma:N:2} and Proposition \ref{thm2:gamma:N:2}, respectively.
From Figs. 2--4 we can see that Proposition \ref{thm:gamma:N:2} induces a more dense stability region than \cite{SolomonFridmanAut13}, but
guarantees a little sparser stability region than Proposition \ref{thm2:gamma:N:2}.
Therefore, Figs. 1--4 show that Proposition \ref{thm:gamma:N:2}
improves the results in \cite{SolomonFridmanAut13} and that the conditions can be further enhanced by Proposition \ref{thm2:gamma:N:2}.

Let us now compare the number of scalar decision variables
in the LMIs. The LMIs of \cite{SolomonFridmanAut13} have $\{4 n^2+3n\}_{n=2}=~22$ variables. Proposition \ref{thm:gamma:N:2} in this paper not only possess a fewer number
$\{3n^2+2n\}_{n=2}=16$ of variables
but also lead to less conservative results. % than those obtained in \cite{SolomonFridmanAut13}.
In comparison with Proposition \ref{thm:gamma:N:2}, Proposition \ref{thm2:gamma:N:2} slightly improves the results at the price of
 $\{2.5 n^2+2.5n\}_{n=2}=15$ additional decision variables.
\begin{table}
\tiny
\vspace{-.6cm}
\begin{center}
\caption{Example 1: maximum allowable value of $T$ for different $h$}
\begin{tabular}{|c||c|c|c|c|c|c|c|l|}
\hline
\!\!$[\max  T]\!\! $ \!\!   & $10^{-5}$ & 0.01 & 0.15 & 0.34 & 0.35 & 0.36 & Decision\!\! \\
${\large \backslash}$ $h$& &  &  &  &  &  & variables\!\! \\
\hline \cite{SolomonFridmanAut13} & 0.274 & 0.265 & 0.141 & 0.005 & - & - & $22$ \\
\hline Prop. \ref{thm:gamma:N:2} & 0.305 & 0.296 & 0.158 & 0.008 & 0.002 & - & $16$ \\
% % 18-gamma//Gamma_N_2_single_remove_additional_G_3_terms.m %%%%%%%%%%  Kun, use no_extended single
 \hline Prop. \ref{thm2:gamma:N:2} & 0.322 & 0.312 & 0.168 & 0.014 & 0.008 & 0.003 & $31$\\
 \hline
\end{tabular}
\end{center}
\end{table}
\begin{figure}[!htb]
\centering
%\footnotesize
\includegraphics[width=6cm]{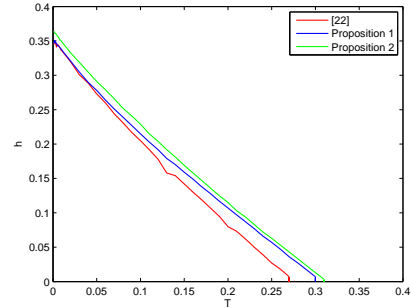}
 \caption{Example 1: tradeoff curve between maximal allowable $T$ and $h$ for Propositions \ref{thm:gamma:N:2} and \ref{thm2:gamma:N:2} compared with the result of \cite{SolomonFridmanAut13} }
\end{figure}
\begin{figure}[!htb]
\centering
%\footnotesize
\includegraphics[width=6cm]{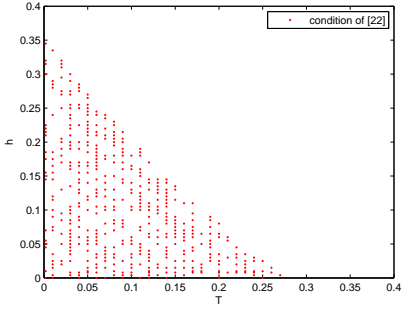}
\caption{Example 1: stability region by the condition of \cite{SolomonFridmanAut13} }
\end{figure}
\begin{figure}[!htb]
\centering
%\footnotesize
\includegraphics[width=6cm]{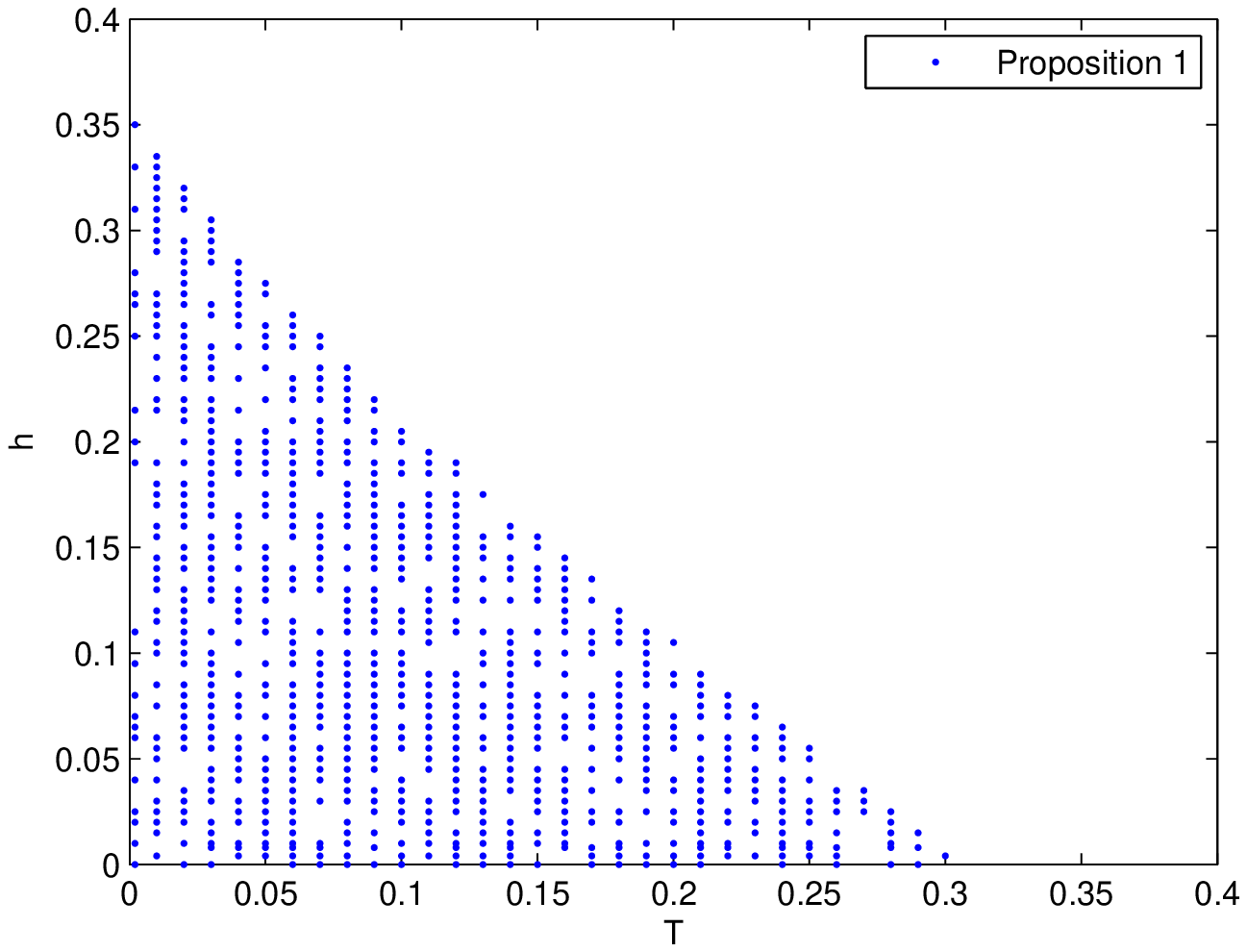}
\caption{Example 1: stability region by Proposition \ref{thm:gamma:N:2}}
\end{figure}
\begin{figure}[!htb]
\centering
%\footnotesize
\includegraphics[width=6cm]{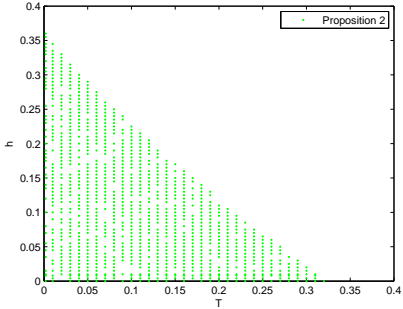}
 \caption{Example 1: stability region by Proposition \ref{thm2:gamma:N:2}}
\end{figure}

%\section{Generalized Jensen inequality with infinite sequences: application to stability analysis of systems with poisson-distributed delays}

\section{Extended Jensen summation inequalities with infinite sequences} \label{sec:infinite:extend:dis}
The objective of this section is to present the discrete counterpart of the
results obtained in Section \ref{sec:infinite:extend}
and to provide extended Jensen summation inequality with infinite sequences.
%The results of this section rely on the following lemma:
We first introduce the following lemma for the discrete counterpart of the integral inequalities (\ref{Inequality:infinite:EF}) and (\ref{Inequality:infinite:doub:EF}):
\begin{lemma}
Assume that there exist an $n \times n$ matrix $R \succ 0,$ a scalar function $M(i) \in {\bbR}^{+}$ and a vector function $x(i) \in {\bbR}^{n}$
%Assume that $\sum_{i=0}^{+\infty} |M(i)| \leq \sum_{i=0}^{+\infty} i|M(i)| <+\infty$
such that the series concerned are convergent. Then the inequality  %(see e.g. \cite{LiuYR08}, \cite{WuFuzzy12}):
\begin{equation}\label{Inequality:infinite:dis:no:extend}
\begin{array}{lll}
\sum_{i=0}^{+\infty} M(i) x^T(i) R x(i)  \\
\hspace{.5cm}\geq M_0^{-1}\Big[\sum_{i=0}^{+\infty} M(i)x(i)\Big]^T R \Big[\sum_{i=0}^{+\infty} M(i)x(i)\Big],
\end{array}
\end{equation}
and its double summation extension
\begin{equation}\label{Inequality:infinite:doub:dis:no:extend}
\begin{array}{lll}
\sum_{i=0}^{+\infty} \sum_{j=k-i-h}^{k-1} M(i) x^T(j) R x(j) \\
\hspace{.5cm}\geq M_{1h}^{-1}
\Big[\sum_{i=0}^{+\infty} \sum_{j=k-i-h}^{k-1} M(i)x(j)\Big]^T R\\
\hspace{1cm}\times \Big[\sum_{i=0}^{+\infty} \sum_{j=k-i-h}^{k-1} M(i)x(j)\Big],
\end{array}
\end{equation}
hold, where
\begin{equation} \label{Notations:M0:dis}
\begin{array}{lll}
 M_0=\sum_{i=0}^{+\infty} M(i),\\
 M_{1h}=\sum_{i=0}^{+\infty} (i+h) M(i).
\end{array}
\end{equation}
%\hspace{16cm}$\blacksquare$
\end{lemma}
The proof of (\ref{Inequality:infinite:dis:no:extend}) and (\ref{Inequality:infinite:doub:dis:no:extend}) follows from \cite{SolomonFridmanAut13} by using sums instead of integrals and  is therefore omitted here. By applying the arguments of Theorem \ref{thm:Inequality:infinite} to the discrete case, we obtain the following theorem for the extended Jensen summation inequality with infinite sequences. Note that this result includes
(\ref{Inequality:infinite:dis:no:extend}) as a special case and that the generalization of (\ref{Inequality:infinite:doub:dis:no:extend})
can be done by the same approach as exploited in Theorem \ref{thm:Inequality:infinite:doub}.
\begin{theorem}
\label{thm:Inequality:infinite:dis}
For a given $n \times n$ matrix $R \succ 0,$ scalar functions $M(i):$ $\bbZ^{+} \to {\bbR}^{+},$ $g(i):$ $\bbZ^{+} \to {\bbR}$ and a vector function $x(i):$ $\bbZ^{+} \to{\bbR}^{n},$
assume that the series concerned are well
defined and  $\sum_{i=0}^{+\infty} M(i) g(i)=0$ with $g(i)$ not identically zero. Then the following inequality holds:
\begin{equation}
\label{Inequality:infinite:dis}
\begin{array}{ll}
\sum_{i=0}^{+\infty} M(i) x^T(i) R x(i)\\
\hspace{.4cm}\geq M_0^{-1}\Big[\sum_{i=0}^{+\infty} M(i)x(i)\Big]^T R \Big[\sum_{i=0}^{+\infty} M(i)x(i)\Big]\\
\hspace{.8cm}+\Big[\sum_{i=0}^{+\infty} M(i) g^2(i) \Big]^{-1}
\Pi^T R \Pi
\end{array}
\end{equation}
with $M_0$ given by (\ref{Notations:M0:dis}) and
\begin{equation}\label{Notations1:Inequality:infinite:dis}
\begin{array}{ll}
% M_0=\sum_{i=0}^{+\infty} |M(i)|,\\
 \Pi=\sum_{i=0}^{+\infty} M(i)g(i)x(i).
\end{array}
\end{equation}
%and
%\begin{equation}\label{Notations2:Inequality:infinite}
%\begin{array}{ll}
% \int_{0}^{+\infty} |K(s)| g(s) ds \equiv 0.
%\end{array}
%\end{equation}
\end{theorem}
{\bf Proof}: See Appendix B.

%{\bf Proof:}
%\hspace{16cm}$\blacksquare$\\
%
%%%%%%%%%%%%%%%%%%%%%%%%%%%%%%%%%%%%%%%%
%%%%%%%%%%%%%%%%%%%%%%%%%%%%%%%%%%%%%%%%
%
In order to apply Theorem \ref{thm:Inequality:infinite:dis} to the stability analysis of time-delay systems,
we take %employing (\ref{Notations2:Inequality:infinite})
\begin{equation}\label{Inequality:infinite:g:dis}
\begin{array}{ll}
g(v)=c(M_0v-M_1), \ \ c \in \bbR \backslash \{0\}, \ \ v \in \bbZ^{+},
\end{array}
\end{equation}
such that $\sum_{i=0}^{+\infty} M(i) g(i)=0$ is satisfied, where
\begin{equation}\label{Inequality:infinite:M1:dis}
\begin{array}{ll}
M_1=\sum_{i=0}^{+\infty} iM(i).
\end{array}
\end{equation}
Hence, we have
\begin{equation}\label{Inequality:infinite:g:2:dis}
\begin{array}{ll}
 \sum_{i=0}^{+\infty} M(i) g^2(i)=c^2\sum_{i=0}^{+\infty} M(i) (M_0i-M_1)^2\\
 \hspace{2.7cm}=c^2(M_0^2M_2-M_0M_1^2),\\
 \Pi=\sum_{i=0}^{+\infty} M(i) g(i)  x(i) \\
 \hspace{.4cm}=c\Big[M_0\sum_{i=0}^{+\infty} iM(i) x(i) -M_1\sum_{i=0}^{+\infty} M(i) x(i) \Big],
\end{array}
\end{equation}
where
\begin{equation}\label{Inequality:infinite:M2:dis}
\begin{array}{ll}
M_2=\sum_{i=0}^{+\infty} i^2M(i).
\end{array}
\end{equation}
%%%%%%%%%%%%%%%%%%%%%%%%%%%%%%%%%%%%%%%%%%%%
%%%%%%%%%%%%%%%%%%%%%%%%%%%%%%%%%%%%%%%%%%%%
From (\ref{Inequality:infinite:g:dis}) and (\ref{Inequality:infinite:g:2:dis}), Theorem \ref{thm:Inequality:infinite:dis} is reduced to the following corollary, which will be employed in the next section for the stability analysis of discrete-time systems with poisson-distributed delays.
%%%%%%%%%%%%%%%%%%%%%%%%%%%%%%%%%%%%%%
\begin{corollary}
\label{Cor:infinite:dis}
Given an $n \times n$ matrix $R \succ 0,$ a scalar function $M(i):$ $\bbZ^{+} \to {\bbR}^{+}$ and a vector function
$x(i):$ $\bbZ^{+} \to{\bbR}^{n}$
such that the series concerned are well
defined, the following inequality holds:
\begin{equation}
\label{Inequality:infinite:cor:dis}
\begin{array}{ll}
\sum_{i=0}^{+\infty} M(i) x^T(i) R x(i)\\
\hspace{.4cm}\geq M_0^{-1}\Big[\sum_{i=0}^{+\infty} M(i)x(i)\Big]^T R \Big[\sum_{i=0}^{+\infty} M(i)x(i)\Big]\\
\hspace{.8cm}+\Big(M_2-{M_1^2 \over M_0} \Big)^{-1}
\tilde \Pi^T R \tilde \Pi,
\end{array}
\end{equation}
where $M_0,$ $M_1$ and $M_2$ are given by (\ref{Notations:M0:dis}), (\ref{Inequality:infinite:M1:dis}) and (\ref{Inequality:infinite:M2:dis}), respectively, and
\begin{equation} \nonumber
\begin{array}{ll}
%
%K_2=\int_{0}^{+\infty} s^2|K(s)| ds,\\
\tilde \Pi={M_1 \over M_0}\sum_{i=0}^{+\infty} M(i) x(i)-\sum_{i=0}^{+\infty} iM(i) x(i).
\end{array}
\end{equation}
\end{corollary}
%\hspace{16cm}$\blacksquare$
%
%\subsection{Discrete-time systems with poisson-distributed delays}
%The efficiency of extended Jensen summation inequality
%will be demonstrated through the stability analysis of linear discrete-time systems with poisson-distributed delays.
\section{Stability analysis of discrete-time systems with poisson-distributed delays} \label{sec:sys:poisson}
In this section, we will demonstrate the efficiency of the extended Jensen summation inequality (\ref{Inequality:infinite:cor:dis})
through the stability analysis of linear discrete-time systems with poisson-distributed delays.
Consider the following system:
\begin{equation}
\label{sys:dis}
\begin{array}{lll}
x(k+1)=Ax(k)+ A_1 \sum_{\tau=0}^{+\infty} p(\tau)x(k-\tau), \ \ k \in {\bbZ}^{+},
\end{array}
\end{equation}
where $x(k)\in {\bbR}^{n}$ is the state vector, the system matrices $A$ and $A_1$  are constant  with
appropriate dimensions. We do not allow $A$ and $A+A_1$ to be Schur stable.
The initial condition is given as
$\col \{x(0), x(-1), x(-2), \dots \}=\col \{\phi(0), \phi(-1), \phi(-2), \dots \}.$
The function $p(v),$ $v \in {\bbZ}^{+},$ is a poisson distribution with a fixed time gap $h \in {\bbZ}^{+}$:
\begin{equation}
\begin{array}{lll}  \nonumber %\label{poisson:delay:dis}
p(v)=\left\{\begin{array} {cl}
 \frac{e^{-\lambda}\lambda^{v-h}}{(v-h)!}  & v \geq h,\\
0 & v< h,
\end{array} \right.
\end{array}
\end{equation}
where $\lambda >0$ is a parameter of the distribution.
The mean value of $p$ is $h+\lambda$.
Due to the fact that
\begin{equation} \nonumber
\begin{array}{lll}
\sum_{\tau=0}^{+\infty} p(\tau)x(k-\tau)&=&\sum_{\tau=h}^{+\infty} p(\tau)x(k-\tau)\\
&=&\sum_{\tet=0}^{+\infty} p(\tet+h)x(k-\tet-h),
\end{array}
\end{equation}
we arrive at the equivalent system:
\begin{equation}
\label{clc:dis}
\begin{array}{lll}
x(k+1)\!=\!Ax(k)\!+\!A_1 \sum_{\tau=0}^{+\infty} {\rm P}(\tau)x(k-\!\tau\!-h), \ k \in {\bbZ}^{+},
\end{array}
\end{equation}
where ${\rm P}(\tau)=\frac{e^{-\lambda}\lambda^{\tau}}{\tau!} $.
%The system (\ref{clc:dis}) is said to be exponentially stable with the decay rate $\delta$
%if there exist constants $ b >0,$ $0<\delta<1$  such that
%%
%$
%|x(k)|^2\leq b \delta^{2k} \|\phi \|_c^2, \
%\forall k\geq 0, \ k \in \bbN,
%$
%holds, where $\|\phi \|_c=\sup_{s \leq 0, s \in \bbZ} |\phi(s)|$.
We next derive LMI conditions for the
asymptotic stability of (\ref{clc:dis}) via a
direct Lyapunov method.

Denoting
\begin{equation} \nonumber
\begin{array}{lll}
f(k)=\sum_{\tau=0}^{+\infty} {\rm P}(\tau)x(k-\tau-h), \ \ k \in {\bbZ}^{+},
\end{array}
\end{equation}
the system (\ref{clc:dis}) can be transformed into the following augmented form
\begin{equation}
\label{clc:dis:aug}
\begin{array}{lll}
x(k+1)=Ax(k)+A_1 f(k), \\
f(k+1)= \sum_{\tau=0}^{+\infty} \frac{e^{-\lambda}\lambda^{\tau}}{\tau!} x(k+1-\tau-h)\\
\hspace{.3cm}= e^{-\lambda} x(k+1-h)+  \sum_{\tau=1}^{+\infty} \frac{e^{-\lambda}\lambda^{\tau}}{\tau!} x(k+1-\tau-h)\\
\hspace{.3cm}= e^{-\lambda} x(k+1-h)+  \sum_{\tau=0}^{+\infty} \frac{e^{-\lambda}\lambda^{\tau+1}}{(\tau+1)!} x(k-\tau-h)\\
\hspace{.3cm}= e^{-\lambda} Ax(k-h)+e^{-\lambda} A_1 f(k-h)\\
\hspace{.5cm}+  \sum_{\tau=0}^{+\infty} {\rm Q}(\tau) x(k-\tau-h),
\end{array}
\end{equation}
where ${\rm Q}(\tau)=\frac{e^{-\lambda}\lambda^{\tau+1}}{(\tau+1)!}$.
It is noted that the augmented system (\ref{clc:dis:aug}) has not only distributed but also discrete delays. This is different from augmented system (\ref{clc:aug}) for the case of gamma-distributed delays. %the last two terms $V_{S_1}(k)$ and $V_{S_2}(k)$ are added to ``compensate" the delayed terms
Moreover, we find that
\begin{equation}
\label{sum:pdf:dis}
\begin{array}{lll}
 \sum_{i=0}^{+\infty}(i+h){\rm P}(i)=\lambda+h,\\
 \sum_{i=0}^{+\infty}{\rm Q}(i)=1-e^{-\lambda},\\
 \sum_{i=0}^{+\infty}i{\rm Q}(i)=\lambda-1+e^{-\lambda}\bydef {\bar {\rm Q}}_1,\\
 \sum_{i=0}^{+\infty}i^2{\rm Q}(i)=\lambda^2-\lambda+1-e^{-\lambda}\bydef {\bar {\rm Q}}_2,\\
 \sum_{i=0}^{+\infty}(i+h){\rm Q}(i)=\lambda\!+\!(1-e^{-\lambda})(h\!-\!1) \bydef {\bar {\rm Q}}_{1h}.
\end{array}
\end{equation}
Consider system (\ref{clc:dis:aug}) with both distributed and discrete delays. The stability analysis will be based on the following discrete-time LKF:
\begin{equation} \nonumber
\begin{array}{lll}
V(k)\!=\!{\hat x}^T(k) \hat W \hat x(k)\!+\! \sum_{i=1}^{2} \Big [V_{G_i}(k)+V_{H_i}(k)+V_{S_i}(k)\Big],\\
V_{G_1}(k)= \sum_{i=0}^{+\infty}\sum_{s=k-i-h}^{k-1}{\rm P}(i) x^T(s) G_1 x(s),\\
V_{H_1}(k)= \sum_{i=0}^{+\infty}\sum_{j=1}^{i+h}\sum_{s=k-j}^{k-1}{\rm P}(i) \eta_1^T(s) H_1 \eta_1(s),\\
V_{G_2}(k)= \sum_{i=0}^{+\infty}\sum_{s=k-i-h}^{k-1}{\rm Q}(i) x^T(s) G_2 x(s),\\
V_{H_2}(k)= \sum_{i=0}^{+\infty}\sum_{j=1}^{i+h}\sum_{s=k-j}^{k-1}{\rm Q}(i) \eta_1^T(s) H_2 \eta_1(s),\\
V_{S_1}(k)= \sum_{s=k-h}^{k-1} x^T(s) S_1x(s)\\
\hspace{1.5cm}+h\sum_{j=-h}^{-1}\sum_{s=k+j}^{k-1} \eta_1^T(s) R_1 \eta_1(s),\\
V_{S_2}(k)= \sum_{s=k-h}^{k-1} f^T(s) S_2f(s)\\
\hspace{1.5cm}+h\sum_{j=-h}^{-1}\sum_{s=k+j}^{k-1} \eta_2^T(s) R_2 \eta_2(s),
\end{array}
\end{equation}
%\begin{equation} \nonumber
%\begin{array}{lll}
%%V(k)\!=\!{\hat x}^T(k) W \hat x(k)\!+\! \sum_{i=1}^{2} \Big [V_{G_i}(k)+V_{H_i}(k)+V_{S_i}(k)\Big],\\
%%V_{G_1}(k)= \sum_{i=0}^{+\infty}\sum_{s=k-i-h}^{k-1}{\rm P}(i) x^T(s) G_1 x(s),\\
%%V_{H_1}(k)= \sum_{i=0}^{+\infty}\sum_{j=1}^{i+h}\sum_{s=k-j}^{k-1}{\rm P}(i) \eta_1^T(s) H_1 \eta_1(s),\\
%%V_{G_2}(k)= \sum_{i=0}^{+\infty}\sum_{s=k-i-h}^{k-1}{\rm Q}(i) x^T(s) G_2 x(s),\\
%%V_{H_2}(k)= \sum_{i=0}^{+\infty}\sum_{j=1}^{i+h}\sum_{s=k-j}^{k-1}{\rm Q}(i) \eta_1^T(s) H_2 \eta_1(s),\\
%V_{S_1}(k)= \sum_{s=k-h}^{k-1} x^T(s) S_1x(s)\\
%\hspace{1.5cm}+h\sum_{j=-h}^{-1}\sum_{s=k+j}^{k-1} \eta_1^T(s) R_1 \eta_1(s),\\
%V_{S_2}(k)= \sum_{s=k-h}^{k-1} f^T(s) S_2f(s)\\
%\hspace{1.5cm}+h\sum_{j=-h}^{-1}\sum_{s=k+j}^{k-1} \eta_2^T(s) R_2 \eta_2(s),
%\end{array}
%\end{equation}
where $\hat W \succ 0,$ $G_i \succ 0,$ $H_i \succ 0,$ $S_i \succ 0,$ $R_i \succ 0,$ $i=1,2,$
%
%\begin{equation}\label{W:dis}
%\begin{array}{lll}
%\hat W=\left[\begin{array} {ccc}
%P & Q\\
%{*} & Z
%\end{array} \right] \succ 0,
%\end{array}
%\end{equation}
and
\begin{equation} \nonumber
\begin{array}{lll}
\hat x(k)=\col\{x(k), \ f(k)\}, \\ %=F_1 \xi(k),\\
%\bar x(k+1)=\col\{x(k+1), f(k+1)\}=F_0 \xi(k),\\
\eta_1(k)=x(k+1)-x(k), \\ %=F_{01} \xi(k),\\
\eta_2(k)=f(k+1)-f(k). %=F_{02} \xi(k).
\end{array}
\end{equation}
Here the last two terms $V_{S_1}(k)$ and $V_{S_2}(k)$ are added to compensate the delayed terms $x(k-h)$ and $f(k-h)$ of (\ref{clc:dis:aug}), respectively. Therefore, for system (\ref{clc:dis:aug}) with $h=0,$ the terms $V_{S_1}(k)$ and $V_{S_2}(k)$ are not necessary.
From standard arguments, we arrive at the following result for the asymptotic stability of (\ref{clc:dis:aug}):
%the condition $\Delta V(k)=V(k+1)-V(k)$
%guarantees the asymptotic stability of (\ref{clc:dis:aug}). Therefore, we arrive at the following result:

\begin{proposition}
\label{thm:poisson:dis:extend}
Given a real scalar $\lambda>0$ and an integer $h \geq 0,$
assume that there exist $2n\times 2n$ positive definite matrix $\hat W$ and $n\times n$ positive definite matrices
$G_i,$ $H_i,$ $S_i,$ $R_i,$ $i=1,2,$ such that the following LMI is satisfied:
\begin{equation} \label{hatPsi:thm:poisson:dis:extend}
\begin{array}{lll}
\hat \Xi&=&\hat \Sigma\!+\!\hat F_0^T \hat W \hat F_0\!-\!\hat F_1^T \hat W \hat F_1-(\lambda+h)^{-1}\hat F_{12}^T H_1 \hat F_{12}\\
&&+\hat F_{01}^T [(\lambda\!+\!h)H_1+{\bar {\rm Q}}_{1h}H_2+h^2R_1]\hat F_{01}\\
&&+ h^2 \hat F_{02}^T R_2 \hat F_{02}-{{\bar {\rm Q}}_{1h}}^{-1}\hat F_{15}^T H_2 \hat F_{15}- \hat F_{13}^T R_1 \hat F_{13}\\
&&- \hat F_{24}^T R_2 \hat F_{24}-\Big(\bar Q_2-{\bar Q_1^2 \over 1-e^{-\lambda}} \Big)^{-1}
\hat F_{25}^T G_2 \hat F_{25} \prec 0,
\end{array}
\end{equation}
where $\bar Q_1,$ $\bar Q_2$ and $\bar Q_{1h}$ are given by (\ref{sum:pdf:dis}), $\hat \Sigma={\rm diag}\{S_1+G_1+(1-e^{-\lambda})G_2, -G_1+S_2, -S_1, -S_2, -(1-e^{-\lambda})^{-1}G_2\}$,
\begin{equation} \nonumber %\label{Notations:thm:poisson:dis:extend}
\begin{array}{lll}
\hat F_0=\left[\begin{array} {ccccc}
A & A_1 & 0 & 0 &0 \\
0 & 0 & e^{-\lambda} A & e^{-\lambda} A_1 &I
\end{array} \right],\\
\hat F_1=\left[\begin{array} {ccccc}
I & 0 & 0 & 0 &0 \\
0 & I & 0 & 0 &0
\end{array} \right],\
\hat F_{01}=[A\!-\!I \ A_1 \ 0 \ 0 \ 0],\\
\hat F_{02}=[0 \ -\!I \ \ e^{-\lambda} A \ \ e^{-\lambda} A_1 \ \ I],\
\hat F_{12}=[I \ -\!I \ 0 \ 0 \ 0],\\
\hat F_{13}=[I \ 0 \ -\!I \ 0 \ 0],\
\hat F_{15}=[(1-e^{-\lambda})I \ 0 \ 0 \ 0 \ -\!I],\\
\hat F_{24}=[0 \ I \ 0 \ -\!I \ 0],\
\hat F_{25}=[0 \ -\lambda I \ 0 \ 0 \  ({\bar Q_1 \over 1-e^{-\lambda}}+1)I].
\end{array}
\end{equation}
Then the system (\ref{clc:dis:aug}) is asymptotically stable. %\hspace{8.5cm}$\blacksquare$
\end{proposition}
{\bf Proof:}
Define
\begin{equation}\nonumber %\label{xi:dis}
\begin{array}{lll}
q(k)= \sum_{\tau=0}^{+\infty} {\rm Q}(\tau) x(k-\tau-h),\\
\xi(k)=\col\{x(k), f(k), x(k-h), f(k-h), q(k)\}.
\end{array}
\end{equation}
By taking difference of $V(k)$ along (\ref{clc:dis:aug}) and applying Jensen inequalities with finite sequences (see e.g., Chapter 6 of \cite{FridmanBook14}), we have
\begin{equation}\label{V1:dis}
\begin{array}{lll}
{\hat x}^T(k+1) \hat W \hat x(k+1)-{\hat x}^T(k)  \hat W \hat x(k)\\
=\xi^T(k)[\hat F_0^T \hat W \hat F_0-\hat F_1^T \hat W \hat F_1]\xi(k)
\end{array}
\end{equation}
and
\begin{equation}
\begin{array}{lll}
 \sum_{i=1}^{2} \Big [V_{G_i}(k+1)+V_{H_i}(k+1)+V_{S_i}(k+1)\\
\hspace{1.2cm}-V_{G_i}(k)-V_{H_i}(k)-V_{S_i}(k)\Big]\\
=\xi^T(k)\Big [\hat \Sigma+\hat F_{01}^T [(\lambda+h)H_1+{\bar {\rm Q}}_{1h}H_2+h^2R_1] \hat F_{01}\\
\hspace{1.4cm}-(\lambda+h)^{-1}\hat F_{12}^T H_1 \hat F_{12}
+h^2 \hat F_{02}^T R_2 \hat F_{02}\\
\hspace{1.4cm}- \hat F_{13}^T R_1 \hat F_{13}- \hat F_{24}^T R_2 \hat F_{24}\Big] \xi(k)\\
\hspace{.5cm}+(1-e^{-\lambda})^{-1}q^T(k)G_2 q(k) \\
\hspace{.5cm} -\sum_{i=0}^{+\infty}{\rm Q}(i) x^T(k-i-h)G_2 x(k-i-h)\\
\hspace{.5cm}  -\sum_{i=0}^{+\infty}\sum_{s=k-i-h}^{k-1}{\rm Q}(i) \eta_1^T(s)H_2 \eta_1(s).
%\hspace{.5cm} -h\sum_{s=k-h}^{k-1}\eta_1^T(s) R_1 \eta_1(s)-h\sum_{s=k-h}^{k-1}\eta_2^T(s) R_2 \eta_2(s).
\end{array}
\end{equation}
Applying the generalized Jensen inequality (\ref{Inequality:infinite:cor:dis}) with infinite sequences, we obtain
\begin{equation}\label{V:G2:poisson:dis:extend}
\begin{array}{lll}
-\sum_{i=0}^{+\infty}{\rm Q}(i) x^T(k-i-h)G_2 x(k-i-h)\\
%\leq -{\bar {\rm Q}}_0^{-1}q^T(k)G_2 q(k)\\
%\quad -\Big(\bar Q_2-{\bar Q_1^2 \over  {\bar {\rm Q}}_0} \Big)^{-1}
%\Big[{\bar Q_1 \over  {\bar {\rm Q}}_0}q(k)-\sum_{i=0}^{+\infty}i{\rm Q}(i) x(k-i-h)\Big]^T G_2 \Big[{\bar Q_1 \over  {\bar {\rm Q}}_0}q(k)-\sum_{i=0}^{+\infty}i{\rm Q}(i) x(k-i-h)\Big]\\
%
\leq-(1-e^{-\lambda})^{-1}q^T(k)G_2 q(k)\\
\quad -\Big(\bar Q_2-{\bar Q_1^2 \over  1-e^{-\lambda}} \Big)^{-1}
\Big[\Big({\bar Q_1 \over  1-e^{-\lambda}}\!+\!1\Big)q(k)\!-\!\lambda f(k)\Big]^T \!G_2\\
\hspace{0.7cm} \times \Big[\Big({\bar Q_1 \over  1-e^{-\lambda}}+1\Big)q(k)-\lambda f(k)\Big]\\
=-(1-e^{-\lambda})^{-1}q^T(k)G_2 q(k)\\
\hspace{0.4cm} -\Big(\bar Q_2-{\bar Q_1^2 \over  1-e^{-\lambda}} \Big)^{-1}
\xi^T(k)\hat F_{25}^T G_2 \hat F_{25}\xi(k).
\end{array}
\end{equation}
%
%where $q(k),$ $\xi(k)$ are given by (\ref{xi:dis}), and
%%
%\begin{equation} \label{Notations:thm:poisson:dis:extend}
%\begin{array}{lll}
%F_{25}=[0 \ -\lambda I \ 0 \ 0 \  ({\bar Q_1 \over \bar Q_0}+1)I].
%\end{array}
%\end{equation}
%
%
%
%
Furthermore, the application of  (\ref{Inequality:infinite:doub:dis:no:extend}) leads to
\begin{equation}\label{V:H2:poisson:dis}
\begin{array}{lll}
 -\sum_{i=0}^{+\infty}\sum_{s=k-i-h}^{k-1}{\rm Q}(i) \eta_1^T(s)H_2 \eta_1(s) \\
\leq \! -\Big[\!\sum_{i=0}^{+\infty}(i\!+\!h){\rm Q}(i)\Big]^{-1}\! \Big[\!\sum_{i=0}^{+\infty}\!\sum_{s=k-i-h}^{k-1}\!{\rm Q}(i) \eta_1(s)\Big]^T\!\!\!H_2 \\
\hspace{.5cm}\times \Big[\sum_{i=0}^{+\infty}\sum_{s=k-i-h}^{k-1}{\rm Q}(i) \eta_1(s)\Big] \\
=-{{\bar {\rm Q}}_{1h}}^{-1}[(1\!-\!e^{-\lambda})x(k)\!-\!q(k)]^TH_2 [(1\!-\!e^{-\lambda})x(k)\!-\!q(k)]\\
= -{{\bar {\rm Q}}_{1h}}^{-1}\xi^T(k)\hat F_{15}^T H_2 \hat F_{15} \xi(k).
\end{array}
\end{equation}
Therefore, (\ref{V1:dis})--(\ref{V:H2:poisson:dis}) yield $\Delta V(k)=V(k+1)-V(k)\leq \xi^T(k) \hat \Xi \xi(k).$  Then if
(\ref{hatPsi:thm:poisson:dis:extend}) holds for given scalars $\lambda>0$ and $h \geq 0$, the system (\ref{clc:dis:aug}) is asymptotically stable. %\hspace{4cm}$\blacksquare$
\begin{remark}
\label{remark:poisson:dis:no:extend}
The LMI condition in Proposition \ref{thm:poisson:dis:extend} is derived by employing the generalized Jensen inequality (\ref{Inequality:infinite:cor:dis}).
The system (\ref{clc:dis:aug}) can be alternatively analyzed
by inequality (\ref{Inequality:infinite:dis:no:extend}).
In this case, (\ref{V:G2:poisson:dis:extend}) is reduced to:
\begin{equation}\nonumber %\label{G2:no:extend:dis}
\begin{array}{lll}
 -\sum_{i=0}^{+\infty}{\rm Q}(i) x^T(k-i-h)G_2 x(k-i-h) \\
\hspace{.3cm} \leq -(1-e^{-\lambda})^{-1}q^T(k)G_2 q(k).
\end{array}
\end{equation}
It yields  $\hat \Xi_{|\hat F_{25}=0} \prec 0$, which is more conservative than the condition proposed in Proposition \ref{thm:poisson:dis:extend}
since the matrix $-\Big(\bar Q_2-{\bar Q_1^2 \over 1-e^{-\lambda}} \Big)^{-1}
\hat F_{25}^T G_2 \hat F_{25}$ of (\ref{hatPsi:thm:poisson:dis:extend})  is negative definite.
%
%with  $\hat \Psi$ given by (\ref{hatPsi:thm:poisson:dis:extend})
\end{remark}

\begin{remark}
Both conditions in Proposition \ref{thm:poisson:dis:extend} and Remark \ref{remark:poisson:dis:no:extend} are derived by the use of inequality (\ref{Inequality:infinite:doub:dis:no:extend}).
It is worth noting that the results could be further improved (in the
$(\lambda, h) $ plane preserving the stability) by the discrete counterpart of Theorem \ref{thm:Inequality:infinite:doub}. %extending inequality (\ref{Inequality:infinite:doub:dis:no:extend}) to a more general one.
\end{remark}

%%%%%%%%%%%%%%%%%%%%%%%%%%%%%%%%
\subsection{Example 2}
%************************ m-files:
%*****************************
% Use For function
%*****************
% 17-possion//possion_general_For.m %%%%%%%%%%  use No_extended single
% 17-possion//possion_general_extend_G1H1_For.m %%%%%%%%%%  use extended single
% 17-possion//possion_general_extend_For.m %%%%%use extended single, No G1, H1
%************************
% No Use For function
%**********************
% 17-possion//possion_general_No_For.m %%%%%%%%%%  use No_extended single
% 17-possion//possion_general_extend_No_For.m %%%%%%%%%%  use extended single
% 17-possion//possion_general_extend_No_For.m %%%%%use extended single, No G1, H1
%**************************
%Plot!!!*****************************
% 17-possion//plot//Plot_possion_general %%%%%%%%%%
% 17-possion//plot//Plot_possion_general_extend_G1H1_For
%
Consider the linear discrete-time system (\ref{sys:dis}) with
\begin{equation} \nonumber
A= \left[
\begin{array}{cc}
       -0.5& 0 \\
       0 & 1
      \end{array}\right]  \ \  \text{and} \ \
A_1= \left[
\begin{array}{cc}
       -0.5& 0.8 \\
       0.5 & -0.2
      \end{array}\right].
\end{equation}
Here neither $A$ nor $A+A_1$ is Schur stable. %Therefore, the delay has stabilizing effect.
For $h=0$ the values of $\lambda$ that guarantee
the asymptotic stability  of the system by Remark \ref{remark:poisson:dis:no:extend} and Proposition \ref{thm:poisson:dis:extend} are shown in Fig. 5, where we can see that the results achieved by Proposition \ref{thm:poisson:dis:extend} are less conservative than those obtained by Remark \ref{remark:poisson:dis:no:extend}. It is noted that Proposition \ref{thm:poisson:dis:extend} and Remark \ref{remark:poisson:dis:no:extend} possess the same number
$\{5n^2+4n\}_{n=2}=28$ of variables.
%Compare now the number of scalar decision variables
%in LMIs of different methods. LMIs of \cite{SolomonFridmanAut13} have the number $4 n^2+3n$ of variables. Note that our conditions in Theorem \ref{thm:gamma:N:2} not only possess a fewer number
%$3n^2+2n$ of variables
%but also lead to less conservative results than those obtained in \cite{SolomonFridmanAut13}.
%Compared with Theorem \ref{thm:gamma:N:2}, Theorem \ref{thm2:gamma:N:2} slightly improves the results at the price of
% $2.5 n^2+2.5n$ more number of decision variables.
%
%\begin{figure}
%  \centering
%  \subfigure[]{
%    %\label{fig:subfig:a} %% label for first subfigure
%    \includegraphics[width=7cm]{possion_general_extend_G1H1.eps}}
%  \hspace{.3in}
%  \subfigure[]{
%    %\label{fig:subfig:b} %% label for second subfigure
%    \includegraphics[width=7cm]{possion_general.eps}}
%  \caption{(a) ;
%(b) .}
%  %\label{fig:subfig} %% label for entire figure
%\end{figure}
%
\begin{figure}[!htb]
\centering
%\footnotesize
\includegraphics[width=6cm]{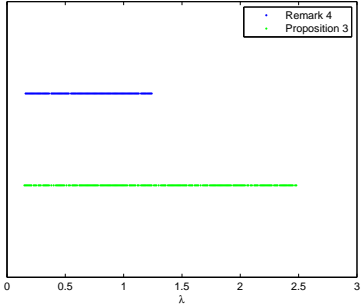}
\caption{Example 2: stabilizing values of $\lambda$}
\end{figure}
%
%\begin{figure}[!htb]
%\centering
%%\footnotesize
%\includegraphics[width=6cm]{possion_general_extend_G1H1}
%\caption{Example 2: stability region by Proposition \ref{thm:poisson:dis:extend}}
%\end{figure}

\section{Conclusions}\label{sec:conclusions}
In this paper, we have
%presented a more general integral inequality with finite intervals of integration.
%Then by the same technique we have
provided extended Jensen integral inequalities. For the discrete counterpart we have generalized Jensen summation inequality.
Applications to the stability analysis of linear continuous-time systems with gamma-distributed delays
and linear discrete-time systems with poisson-distributed delays demonstrated the advantages of these
generalized  inequalities.
In both cases,  the considered infinite distributed delays with a gap have stabilizing effects.
 The
future research may include other applications of these developed inequalities.

%first, by virtue of a more general auxiliary function  we develop an extended integral inequality, which corresponds to finite intervals of integration and includes
%the one proposed in \cite{SeuretAut13wirtinger} as a particular case.
%Then by the same methodology we present generalized integral inequalities and its double integral extension, which are on the infinite intervals of
%integration and can be applied to improve the stability result for linear continuous-time systems with gamma-distributed delays.
%Moreover, for discrete counterpart we provide extended Jensen inequality with infinite sequences, which leads to less conservative stability conditions for linear discrete-time systems with poisson-distributed delays.
%In both continuous-time and discrete-time cases,  the considered infinite distributed delays with a gap in the paper have stabilizing effect. Following \cite{SolomonFridmanAut13}, we derive the results via augmented Lyapunov functionals.
\vspace{-.2cm}
\section*{Appendix A}
{\bf Proof of Theorem \ref{thm:Inequality:infinite:doub}}
%
%For any continuous function $\omega: [t-\tet-h, t] \to \bbR^n$ and any scalar function
%$g: [t-\tet-h, t] \to \bbR$, assume that the integrations concerned are well
%defined.
Following the proof of Theorem~\ref{thm:Inequality:infinite}, we define a function $f(r, u)$ for all
$u \in [0, +\infty)$, $r \in [t-u-h, t]$ by
$
f(r, u)=\hat f(r, u)-\sqrt {K(u)} g(r) \hat \Theta,
$
where $\hat \Theta \in \bbR^n$ is a constant vector to be defined and $\hat f(r, u)=\sqrt {K(u)} \Big [\omega(r)-K_{1h}^{-1} \int_{0}^{+\infty} \int_{t-\tet-h}^{t} K(\tet)  \omega(s) ds d \tet \Big]$ with $K_{1h}$ given by (\ref{Inequality:infinite:doub}). Since $R \succ 0$ we have
\begin{equation}\nonumber %\label{Inequality:infinite:doub:proof:1}
\begin{array}{llll}
 0 &\leq&  \int_{0}^{+\infty} \int_{t-\tet-h}^{t} f^T(s, \tet) R f(s, \tet) ds d \tet\\
 &=&  \int_{0}^{+\infty}\int_{t-\tet-h}^{t} \hat f^T(s, \tet) R \hat f(s, \tet)ds d \tet\\
 &&  + \Big[2 K_{1h}^{-1} \hat \Theta^T R \int_{0}^{+\infty} \int_{t-\tet-h}^{t} K(\tet)  \omega(s) ds d \tet \Big] \\
 && \quad \  \times \int_{0}^{+\infty} \int_{t-\tet-h}^{t}K(\tet) g(s) ds d \tet \\
&&  +\Big [\int_{0}^{+\infty}\int_{t-\tet-h}^{t} K(\tet)  g^2(s) ds d \tet\Big] \hat \Theta^T R \hat \Theta \\
&&  -2 \hat \Theta^T R \int_{0}^{+\infty}\int_{t-\tet-h}^{t}  K(\tet) g(s) \omega(s) ds d \tet,
\end{array}
\end{equation}
Rewriting the last two terms as sum of squares together with $\int_{0}^{+\infty} \int_{t-\tet-h}^{t}K(\tet) g(s) ds d \tet=0$
leads to
\begin{equation}\label{Inequality:infinite:doub:proof:2}
\begin{array}{llll}
\int_{0}^{+\infty} \int_{t-\tet-h}^{t}K(\tet) \omega^T(s) R \omega(s) ds d \tet \\
\hspace{1cm}\geq K_{1h}^{-1} \int_{0}^{+\infty} \int_{t-\tet-h}^{t} K(\tet)  \omega^T(s) ds d \tet R\\
\hspace{1cm}\times \int_{0}^{+\infty}\int_{t-\tet-h}^{t} K(\tet)   \omega(s) ds  d \tet\\
\hspace{0.5cm}+\Big[\int_{0}^{+\infty}\int_{t-\tet-h}^{t} K(\tet) g^2(s)ds d \tet \Big]^{-1}
\Sigma^T R \Sigma\\
\hspace{0.5cm}-\Big [\int_{0}^{+\infty}\int_{t-\tet-h}^{t} K(\tet) g^2(s) ds d \tet\Big] [\hat \Theta-\hat \Upsilon]^T R [\hat \Theta-\hat \Upsilon],
\end{array}
\end{equation}
where $\Sigma$ is given by (\ref{Notations1:Inequality:infinite:doub}) and
\begin{equation} \nonumber
\begin{array}{llll}
\hat \Upsilon=  \Big[\int_{0}^{+\infty}\int_{t-\tet-h}^{t} K(\tet) g^2(s)ds d \tet \Big]^{-1} \\
\hspace{.8cm} \times \int_{0}^{+\infty}\int_{t-\tet-h}^{t}  K(\tet) g(s) \omega(s) ds d \tet.
 \end{array}
\end{equation}
Then the inequality (\ref{Inequality:infinite:doub:proof:2}) with $\hat \Theta=\hat \Upsilon$
%
%Since (\ref{Inequality:infinite:doub:proof:2}) holds independently of the choice of $\Theta$, we choose $\Theta=\tilde \Theta$, which leads to
%the maximum of the right-hand side of (\ref{Inequality:infinite:doub:proof:2}) and then yields
implies (\ref{Inequality:infinite:doub}). This concludes the proof. %\hspace{3cm} $\blacksquare$

\section*{Appendix B}
{\bf Proof of Theorem \ref{thm:Inequality:infinite:dis}}
%{\bf Proof}:
%
%For a given vector function $x(i):$ $\bbZ^{+} \to \bbR^n$ and scalar functions
%$M(i):$ $\bbZ^{+} \to \bbR,$ $g(i):$ $\bbZ^{+} \to \bbR$ with  $\sum_{i=0}^{+\infty} |M(i)| g(i)=0$ and $g(i)$
%not equivalent to $0$ for all $i \in \bbZ^{+},$ assume that the series concerned are well
%defined.
Define a function $f(v)$ for all $v \in \bbZ^{+}$ by
$
f(v)=\tilde f(v)- \sqrt {M(v)} g(v) \tilde \Theta,
$
where $\tilde \Theta \in \bbR^n$ is a constant vector to be defined and $\tilde f(v)=\sqrt {M(v)} \Big [x(v)-M_0^{-1} \sum_{i=0}^{+\infty} M(i)x(i)\Big]$ with $M_0$ given by (\ref{Notations:M0:dis}) . Then since $R \succ 0$ it follows that
\begin{equation} \nonumber %\label{Inequality:infinite:proof:1:dis}
\begin{array}{llll}
 0 &\leq&  \sum_{i=0}^{+\infty} f^T(i) R f(i) \\
 &=& \Big[2 M_0^{-1} \tilde \Theta^T R \sum_{i=0}^{+\infty} M(i) x(i)  \Big]  \sum_{i=0}^{+\infty} M(i) g(i)  \\
 &&  + \sum_{i=0}^{+\infty} \tilde f^T(i) R \tilde f(i)\\
 &&  +\Big [\sum_{i=0}^{+\infty} M(i) g^2(i) \Big] \tilde \Theta^T R \tilde \Theta \\
 && -2 \tilde \Theta^T R \sum_{i=0}^{+\infty} M(i) g(i) x(i).
\end{array}
\end{equation}
Representing the last two terms as sum of squares together with $\sum_{i=0}^{+\infty} M(i) g(i)=0$ yields
\begin{equation}\label{Inequality:infinite:proof:2:dis}
\begin{array}{llll}
\sum_{i=0}^{+\infty} M(i) x^T(i) R x(i)\\
\hspace{.4cm} \geq M_0^{-1}\Big[\sum_{i=0}^{+\infty} M(i)x(i)\Big]^T R \Big[\sum_{i=0}^{+\infty} M(i)x(i)\Big]\\
\hspace{.6cm}+\Big[\sum_{i=0}^{+\infty} M(i) g^2(i) \Big]^{-1}\Pi^T R \Pi\\
 \hspace{.6cm} -\Big[\sum_{i=0}^{+\infty} M(i) g^2(i) \Big] [\tilde \Theta-\tilde \Upsilon]^T R [\tilde \Theta-\tilde \Upsilon],
\end{array}
\end{equation}
where $\Pi$ is given by (\ref{Notations1:Inequality:infinite:dis}) and
\begin{equation} \nonumber
\tilde \Upsilon =  \Big[\sum_{i=0}^{+\infty} M(i) g^2(i) \Big]^{-1} \sum_{i=0}^{+\infty}M(i)  g(i) x(i).
\end{equation}
The choice of $\tilde \Theta=\tilde \Upsilon$ results in
the maximum of the right-hand side of (\ref{Inequality:infinite:proof:2:dis}) and thus (\ref{Inequality:infinite:dis}). This concludes the proof. %\hspace{12cm}$\blacksquare$

\bibliography{bibliography}
\bibliographystyle{plain}% plain, the order cited.  %alpha, order of the name %IEEEtran   %abbrv

\end{document}